\documentclass[11pt,a4paper]{article}
\usepackage{jheppub}
\usepackage{amsthm,bm,mathrsfs,stmaryrd}
\usepackage[table]{xcolor}
\usepackage{arydshln,tabu}
\usepackage{tikz}
\usepackage{ytableau}
\usepackage{slashed}
\usepackage{xcolor}
\usepackage{framed}
\definecolor{shadecolor}{gray}{0.925}

\newcommand{\pl}{{\partial}}

\newcommand{\be}{\begin{equation}}
\newcommand{\ee}{\end{equation}}
\newcommand{\ba}{\begin{eqnarray}}
\newcommand{\ea}{\end{eqnarray}}

\newcommand{\mt}[1]{$\mathop{#1}$}

\newcommand{\sst}{\scriptscriptstyle}

\newcommand{\nn}{\nonumber\\}
\newcommand{\eq}{&=&}

\newcommand{\ett}{\,\overset{\sst\rm TT}{=}\,}

\newcommand*\circled[1]{\tikz[baseline=(char.base)]{
  \node[shape=circle,draw, inner sep=1pt] (char) {#1};}}
\newcommand*\dcircled[1]{\tikz[baseline=(char.base)]{
  \node[shape=circle,fill=black!30, draw,inner sep=1pt] (char) {#1};}}

\def\sideremark#1{\ifvmode\leavevmode\fi\vadjust{\vbox to0pt{\vss
 \hbox to 0pt{\hskip\hsize\hskip1em
 \vbox{\hsize3cm\tiny\raggedright\pretolerance10000
 \noindent #1\hfill}\hss}\vbox to8pt{\vfil}\vss}}}%

\makeatletter
\newcommand{\thickhline}{%
    \noalign {\ifnum 0=`}\fi \hrule height 1pt
    \futurelet \reserved@a \@xhline
}
\newcolumntype{"}{@{\hskip\tabcolsep\vrule width 1pt\hskip\tabcolsep}}
\makeatother

\def\l{\lambda}

\def\m{\mu}
\def\n{\nu}

\def\cD{{\cal D}}

\def\cK{{\cal K}}

\ytableausetup{centertableaux}

\author[a]{Euihun JOUNG}
\author[b,c,d]{and Massimo TARONNA}

\affiliation[a]{Department of Physics and Research Institute of Basic
  Science, \\
	Kyung Hee University,\\ Seoul 02447, Korea}

\affiliation[b]{Department of Physics, Princeton University,\\
Jadwin Hall, Princeton, NJ 08544}

\affiliation[c]{Dipartimento di Fisica ``Ettore Pancini'', Universit\`a degli Studi di Napoli Federico II, \\Monte S. Angelo, Via Cintia, 80126 Napoli, Italy}

\affiliation[d]{INFN, Sezione di Napoli, Monte S. Angelo, Via Cintia, 80126 Napoli, Italy}

\emailAdd{euihun.joung@khu.ac.kr}
\emailAdd{mtaronna@princeton.edu}

\title{\centering A note on higher-order vertices of higher-spin fields \\
in flat and (A)dS space}

\abstract{
In this work we classify homogeneous solutions to the Noether procedure in (A)dS for an arbitrary number of external legs and in general dimensions. We also give a review of the corresponding flat space classification and its relation with the (A)dS result presented here. The role of dimensional dependent identities is also investigated.}

\dedicated{Dedicated to the memory of our friend and colleague Dima Polyakov.}

\begin{document}

\maketitle

\section{Introduction}

In this note we address the problem of classifying $n$-point couplings of massive and massless higher-spin fields in both flat space (where a classification was first analysed in \cite{Taronna:2011kt,Taronna:2017wbx}) and (A)dS space where a complete classification has not been yet available.\footnote{Explicit results at quartic or higher orders are scarce, see e.g. \cite{Polyakov:2010sk,Dempster:2012vw,Bengtsson:2016hss,Roiban:2017iqg,Polyakov:2018bja} for various other efforts.} This endeavour is part of the so called Noether procedure program (see e.g. \cite{Berends:1979kg,Berends:1984rq,Berends:1984wp,Barnich:1993vg,Metsaev:2005ar,Boulanger:2008tg,Manvelyan:2010jr,Sagnotti:2010at,Joung:2011ww,Joung:2012fv} for a few non-exhaustive list of references). One of the main points of our approach is to reformulate the Noether program in a form in which locality is not manifest or built in, pursuing the idea of \cite{Taronna:2011kt}. At the same time, one of the advantages of this approach is that it allows to have some quantitative control on the degree of non-locality of the most general Noether solution, as we shall demonstrate. 
More recently, this type of approach taken in \cite{Sleight:2017pcz}, along similar lines,\footnote{See also \cite{Kessel:2015kna,Skvortsov:2015lja,Taronna:2016ats,Taronna:2016xrm} for other related results at the level of unfolded formulation \cite{Vasiliev:1990en}.} led to the conclusion that HS theories in AdS present a degree of non-locality in the same functional class as exchange amplitudes, similar to their flat-space counterparts. This contradicts the initial expectations and raises an issue on the field-theoretical interpretation of (A)dS theories and in \cite{Sleight:2017pcz} a possible reformulation as topological string theories (see e.g. \cite{Engquist:2005yt}) was suggested.\footnote{
Although the formulation of HS theories as field theories in AdS is currently under debate, it is still possible in principle to formulate a Frobenious-Chern Simons theory on non-commutative 10d space based on the higher-spin algebra \cite{Bonezzi:2016ttk}. The precise relation between such approach and a stringy-AdS/CFT picture is still an open question which in order to be settled would require to properly define admissible boundary conditions in the whole non-commutative space. A first proposal in this direction have been set forth and tested to the linearised level in \cite{Iazeolla:2017dxc,DeFilippi:2019jqq}. It would be important to clarify whether such proposal is valid at the fully non-linear level. In any case, classifying admissible boundary conditions of the Frobenious Chern Simons theory on non-commutative space is an interesting mathematical question which might ultimately shed light on how to relate the Frobenious Chern Simons formulation to a more standard stringy/holographic formulation of higher-spin theories at the interacting level.} Even though locality is a key ingredient of the Noether procedure for which so far no convincing replacement has been successfully proposed, we would like to remark here that it is still useful in principle to study the constraints of gauge invariance on $n$-point couplings in (A)dS regardless of their locality properties. Such tensorial structures indeed would still turn out to be holographic dual of corresponding $n$-point spinning structures which satisfy appropriate conservation conditions.\footnote{For $n=3$ this map was written down in \cite{Sleight:2017fpc} for arbitrary integer spinning legs} Such structures are expected to play a role to describe stress-tensor and current correlators and the result presented in this paper would therefore find potentially interesting applications in this context.

In the following, we first give a self-contained introduction of the main concepts of the Noether procedure program following \cite{Joung:2012fv,Taronna:2012gb}, but leaving out for brevity most of the technical details which can be found in the literature.

The starting point is the expansion of the sought action 
and gauge transformations in powers of the fields:
\be
	S=S^{\sst (2)}+S^{\sst (3)}+\cdots\,,\qquad
	\delta_{\varepsilon}\,\varphi=\delta^{\sst (0)}_{\varepsilon}\varphi
	+\delta^{\sst (1)}_{\varepsilon}\varphi+\cdots\,.
\ee
Here, the superscript $(n)$ means that the corresponding term involves $n$-th powers of the fields $\varphi$.
In this weak-field expansion scheme, the gauge invariance of the action is recast into an infinite number of coupled equations:
\be
	\delta_{\varepsilon}\,S=0
	\quad\Rightarrow\quad
	\left\{\begin{array}{cc}
	\delta^{\sst (0)}_{\varepsilon}\,S^{\sst (2)}=0 &\quad \circled{\scriptsize 0}\\
	\delta^{\sst (0)}_{\varepsilon}\,S^{\sst (3)}
	+\delta^{\sst (1)}_{\varepsilon}\,S^{\sst (0)}=0 & 
	\quad  \circled{\scriptsize 1}\\
	\delta^{\sst (0)}_{\varepsilon}\,S^{\sst (4)}
	+\delta^{\sst (1)}_{\varepsilon}\,S^{\sst (3)}
	+\delta^{\sst (2)}_{\varepsilon}\,S^{\sst (2)}=0 & \quad 
	\circled{\scriptsize 2}\\
	\vdots
	\end{array}
	\right.,
	\label{gauge inv}
\ee
where $S^{\sst (2)}$ and $\delta^{\sst (0)}_{\varepsilon}\varphi$ are the free action and the corresponding gauge transformations, respectively.

Higher order parts of the action, $S^{\sst (n\ge3)}$\,, and the gauge transformation, $\delta^{\sst (n\ge1)}_{\varepsilon}\varphi$\,, can be identified starting from $S^{\sst (2)}$ and $\delta^{\sst (0)}_{\varepsilon}\varphi$ by solving the above equations. The strategy is to solve the equations $\circled{\tiny n}$ of \eqref{gauge inv} in two steps as
\be
	S^{\sst (2)}\,,\, \delta^{\sst (0)}_{\varepsilon}\varphi
	\ \ \overset{\dcircled{\tiny 1}}{\bm\longrightarrow} \ \ 
	S^{\sst (3)}
	\ \ \overset{\circled{\tiny 1}}{\bm\longrightarrow} \ \
	\delta^{\sst (1)}_{\varepsilon}\varphi
	\ \ \overset{\dcircled{\tiny 2}}{\bm\longrightarrow} \ \
	S^{\sst (4)}
	\ \ \overset{\circled{\tiny 2}}{\bm\longrightarrow} \ \
	\delta^{\sst (2)}_{\varepsilon}\varphi
	\ \ {\bm\longrightarrow} \ \
	\cdots\,,
\ee
where $\dcircled{\tiny n}$ represents the same condition $\circled{\tiny n}$ but solved on the shell of \emph{free} EoM. In particular, at each order one can first solve for $S^{\sst (n+2)}$ using  $\dcircled{\tiny n}$ and then read off $\delta^{\sst (n)}_{\varepsilon}\varphi$ from $\circled{\tiny n}$\,.
On the other hand, the full non-linear gauge transformations must form an (open) algebra: 
\be
	\delta_{\varepsilon_{1}}\,\delta_{\varepsilon_{2}}\,\varphi
	-\delta_{\varepsilon_{2}}\,\delta_{\varepsilon_{1}}\,\varphi
	\approx \delta_{[\![\varepsilon_{1},\varepsilon_{2}]\!]}\,\varphi\,,
\ee
where the commutator $[\![\varepsilon_{1},\varepsilon_{2}]\!]$ is in principle field-dependent and can be expanded as
\be
	[\![\,\varepsilon_{1}\,,\,\varepsilon_{2}\,]\!]
	=[\![\,\varepsilon_{1}\,,\,\varepsilon_{2}\,]\!]^{\sst (0)}+
	[\![\,\varepsilon_{1}\,,\,\varepsilon_{2}\,]\!]^{\sst (1)}+\cdots\,.
\ee
For the purpose of the present note, 
one first focuses on the lowest-order part of the commutator:
\be
	\delta^{\sst (0)}_{\varepsilon_{1}}\,
	\delta^{\sst (1)}_{\varepsilon_{2}}\,\varphi
	-\delta^{\sst (0)}_{\varepsilon_{2}}\,\delta^{\sst (1)}_{\varepsilon_{1}}\,
	\varphi
	=\delta^{\sst (0)}_{[\![\varepsilon_{1},\varepsilon_{2}]\!]^{(0)}}\varphi\,.
\ee
which is field-independent and can be entirely obtained from $\delta^{\sst (1)}_{\varepsilon}$\, as shown in \cite{Joung:2013nma}.
To sum up, once consistent cubic interactions are determined for a given free theory, then they induce deformations of the gauge transformations and also of the gauge algebra: 
\be
	S^{\sst (3)}\quad \Rightarrow \quad \delta^{\sst (1)}_{\varepsilon}\varphi
	\quad \Rightarrow
	\quad [\![\,\varepsilon_{1}\,,\,\varepsilon_{2}\,]\!]^{\sst (0)}\,.
\ee
Closure then uniquely identify the higher-spin algebras which in turn fixes all coupling constants at cubic order \cite{Sleight:2016dba,Sleight:2016xqq}.
In this sense the free Fronsdal action contains all the information about the deformation of the gauge symmetries of the theory and their representations which together with closure fully specifies the higher-spin algebra structure constants.
In the following, we shall consider massless, bosonic, symmetric fields in both flat and (A)dS space.  

\subsubsection*{Ambient-space formulation and Transverse and Traceless part}

In order to treat fields in (A)dS in a simple manner, we introduce the ambient-space formulation where fields $\varphi_{\mu_{1}\cdots\mu_{s}}(x)$ are described through the corresponding ambient avatars $\Phi_{\sst M_{1}\cdots M_{s}}(X)$, defined in a $(d+1)$-dimensional flat space and subject to homogeneity and tangentiality conditions:
\be
	(X\cdot\partial_{X}-U\cdot\partial_{U}+2-\mu)\,\Phi(X,U)=0\,,
	\qquad
	X\cdot \partial_{U}\,\Phi(X,U)=0\,.
	\label{HT}
\ee
The mass of the field is here parametrised by the degree of homogeneity $\mu$\,,
and when the field is massless (that is $\mu=0$), it admits gauge symmetries:
\be
	\delta^{\sst (0)}_{\sst E}\,\Phi
	=U\cdot \partial_{X}\,E
	\qquad
	\big[\,\partial_{U}^{2}\,E=0\,,\ 
	(X\cdot\partial_{X}-U\cdot\partial_{U})E=0\,,
	\ 
	X\cdot\partial_{U}\,E=0\,\big]\,.
	\label{g tr}
\ee
In the following we shall work with ambient space fields satisfying the latter homogeneity conditions. These homogeneity conditions should be contrasted with those used in \cite{Sleight:2016dba,Sleight:2017fpc,Sleight:2017cax} which are more directly tuned to AdS/CFT calculations. The dictionary between the two formulation is however straightforward as discussed in the latter references.
In constructing gauge-invariant interaction vertices, we focus for simplicity on the transverse and traceless (TT) part of the latter disregarding the terms proportional to divergences and traces of the fields. This is equivalent to consider, instead of the full vertices and their gauge variations, their quotient modulo the following equivalence relations:  
\be
	\partial_{U}\!\cdot \partial_{X}\,\Phi \ett 0\,,
	\qquad
	\partial_{U}^{\,2}\,\Phi \ett 0\,;
	\qquad
	\partial_{U}\!\cdot \partial_{X}\,E \ett 0\,,
	\qquad
	\partial_{X}^{\,2}\,E \ett 0\,.
	\label{TT}
\ee
In this setting, the free action assumes a general form and is given simply by
\be
	S^{\sst (2)}[\Phi]\ett -\,\frac12\
	\int_{\rm\sst (A)dS}
	e^{\partial_{U_{1}}\!\cdot\,\partial_{U_{2}}}\,
	\Phi(X,U_{1})\,\partial_{X}^{\,2}\,\Phi(X,U_{2})\,
	\Big|_{\sst U_{i}=0}\,.
	\label{amb free}
\ee
Within this description one can also describe the interacting part of action $S^{\sst (n\ge 3)}[\Phi]$ together with the corresponding gauge transformations $\delta^{\sst (n\ge1)}_{\sst E}\,\Phi$. For instance, the interaction parts of the action can be conveniently expressed as
\be
	S^{\sst (n)}[\Phi] \ett \int_{\sst\rm (A)dS}
	C^{\sst (n)}\ \Phi(X_{1},U_{1})\,\cdots\,\Phi(X_{n},U_{n})\,\Big|_{{}^{X_{i}=X}_{U_{i}=0}}\,,
	\label{n-th order}
\ee
in terms of a differential operator $C^{\sst(n)}$ in $X_{i}$ and $U_{i}$\,. Each operator $C^{\sst(n)}$ is constrained by the gauge-invariance conditions \eqref{gauge inv}.

\paragraph{$H$ vs $G$ structures}

\noindent An important lesson that was first drawn in \cite{Joung:2012rv,Joung:2012hz,Joung:2013nma} is that cubic interactions corresponding to \emph{trivial} deformations of the gauge symmetries are related to the existence of a tensor structure $H_{ij}$\,:
\be
	H_{ij}
	=\partial_{U_{i}}\!\cdot\partial_{X_{j}}\,
	\partial_{U_{j}}\!\cdot\partial_{X_{i}}
	-\partial_{X_{i}}\!\cdot\partial_{X_{j}}\,\partial_{U_{i}}\!\cdot\partial_{U_{j}}\,.
	\label{H}
\ee
The $H_{ij}$'s are operators taking the ambient space curls of the $i$-th and $j$-th fields and contracting them. They are gauge invariant without making use of the on-shell condition, and hence, they do not lead to any deformation of the gauge transformations.

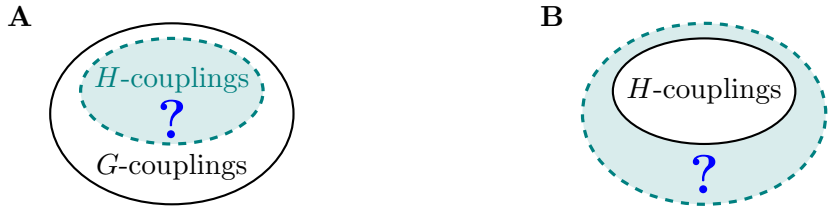
\begin{figure}[h]
\centering
\begin{tikzpicture}
\draw [thick] (0.5,-0.3)  ellipse (1.6cm and 1.2cm);
\node at (0.5,-1) {$G$-couplings};
\draw [teal,dashed,fill=teal!15,very thick] (0.5,0)  ellipse (1.2cm and 0.7cm);
\node at (0.5,0.15) {\color{teal} $H$-couplings};
\node at (0.5,-0.37) {\color{blue} \huge\bf ?};
\draw [teal,fill=teal!15,dashed,very thick] (7.5,-0.3)  ellipse (1.6cm and 1.2cm);
\draw [fill=white,thick] (7.5,0) ellipse (1.2cm and 0.7cm);
\node at (7.5,0) {$H$-couplings};
\node at (7.5,-1.1) {\color{blue} \huge\bf ?};
\node at (-1.5,1) {\bf A};
\node at (5.5,1) {\bf B};
\end{tikzpicture}
\caption{Schematic diagrammatic representation of gauge invariant couplings. The smaller subset represent those couplings which up to a field redefinition are gauge invariant off-shell. The bigger set represent instead the space of couplings whose gauge variation is proportional to the EoMs. It is important to note that there might in principle exist couplings which are not proportional to the EoMs and whose gauge transformation instead is.}
\label{diagram}
\end{figure}

The above $H$-structures should be confronted with the actual classification of massless interactions which can be expressed in terms of what was referred to as $G$-structure in \cite{Sagnotti:2010at}:
\begin{equation}\label{Gstr}
    G=\partial_{U_1}\cdot\partial_{X_2}\,\partial_{U_2}\cdot\partial_{U_3}+\text{cyclic},
\end{equation}
and are in particular proportional to powers $G^n$ of the YM structure $G$ up to appropriately tuned lower derivative terms (see e.g. eq.~(2.4) of \cite{Joung:2013nma}). In particular, in \cite{Joung:2011ww} the above contractions of $\pl_U$ and $\pl_X$ used to define cubic couplings via a generating function notation were defined as
\begin{align}
    Y_i&=\pl_{U_i}\cdot\pl_{X_{i+1}}\,,& Z_i&=\pl_{U_{i-1}}\cdot \pl_{U_{i+1}}
\end{align}
with $i\sim i+3$.
Therefore, the classification of gauge deformations boils down to the question whether a $G$-coupling of the form:
\begin{equation}
    C^{(3)}\sim Y_1^{s_1-k}Y_2^{s_2-k}Y_3^{s_3-k}G^k+\mathcal{O}(\Lambda)\,,
\end{equation}
is expressible (up to field redefinitions) as a $H$-coupling namely a function of the $H_{ij}$ structures \eqref{H} which are off-shell gauge invariant: see e.g. the diagram {\bf A} of Figure \ref{diagram}. On the other hand, one can also view this problem in the opposite perspective, as the question whether there are additional couplings besides those which are trivially gauge-invariant and expressed in terms of the $H$-structures: see the diagram {\bf B} of Figure \ref{diagram}.
The latter viewpoint is natural in the sense that 
$H$-couplings always provide consistent interactions
while $G$-couplings are gauge-invariant on-shell and therefore arise only in some special cases
as, for instance, the case of three-massless-field interactions.
This question was addressed in great detail in \cite{Joung:2012rv,Joung:2012hz}, where it was shown that also in more general classes of gauge interactions (e.g. involving partially-massless or massive fields), 
$G$-couplings arises only when the field masses satisfy specific conditions
whereas $H$-couplings exist always and are furthermore manifestly gauge-invariant off-shell. In this note we shall demonstrate how this simple picture turns out to be applicable also beyond the cubic order and extends to higher-order interactions in a suggestive way.

\paragraph{Summary of higher-order results}

In order to extend the cubic analysis to higher orders, following \cite{Taronna:2011kt}, the main idea is to split the interaction term into two parts:
\be
	C^{\sst (n)}=C^{\sst (n)}_{p}+C^{\sst (n)}_{h}\,.
\ee
Here $C^{\sst (n)}_{p}$ is the particular-solution to $\circled{\scriptsize n}$ in the sense that it deals with the compatibility with the lower $n$ interactions $C^{\sst (m<n)}_{p}$, and $C^{\sst (n)}_{h}$ is any homogeneous-solution part solving $\delta^{\sst (0)}_{\varepsilon}S^{\sst (n)}_{h}+\delta_\varepsilon^{\sst (n)}S^{\sst (2)}=0$ and talking (a priori) with the quadratic action only.
The main point is that, although both $C^{\sst (n)}_{p}$ and $C^{\sst (n)}_{h}$ taken by themselves are generically non-local objects, they admit a simple physical interpretation which allows to find a formal solution to the Noether procedure to any order in a weak field expansion \cite{Taronna:2011kt,Sleight:2016xqq,Sleight:2017pcz}. 
Then, given a certain particular solution $C^{\sst (n)}_{p}$ which we shall discuss in \S\ref{sec:partSol}, one can add any homogeneous solution $C^{\sst (n)}_{h}$. Locality is reformulated in this language as the condition that all non-localities in $C^{\sst (n)}_{p}$ are compensated by a certain choice of $C^{\sst (n)}_{h}$ (if this choice exists). This condition, when a solution exists, recursively relates higher-order contact interactions to lower ones reconstructing a full non-linear theory.

The general solution for $C^{\sst (n)}_{h}$ can now be
obtained by requiring its on-shell gauge invariance w.r.t. the free EoM. Hence, this corresponds to a straightforward generalisation of the cubic-interaction problem to higher orders.\footnote{This problem is equivalent to identifying all deformations that start at the $n$-th order in the fields:
\be
S=S^{\sst (2)}+S^{\sst (n)}+S^{\sst (n+1)}+\cdots\,.
\ee} Its general solution will be shown in this note to be given by
\be
	C^{\sst (n)}_{h}=
	K^{\sst (n)}(\,W^{ij}, H_{ij},H^{jk}_i\,)\,,
	\qquad\qquad [n\ge 4]\,,
\ee
where we also introduced the structure $W^{ij}=\partial_{X_{i}}\!\cdot \partial_{X_{j}}$, which is non-trivial for $n>3$, the $H_{ij}$'s are given in \eqref{H} while $H^{jk}_{i}$'s are:
\be
	H^{jk}_{i}=\partial_{X_{i}}\!\cdot\partial_{X_{j}}\,
	\partial_{X_{k}}\!\cdot\partial_{U_{i}}
	-\partial_{X_{i}}\!\cdot\partial_{X_{k}}\,
	\partial_{X_{j}}\!\cdot\partial_{U_{i}}\,.
\ee
The $H$ structures above are nothing but the generalisations of the cubic $H$ structures to the case in which more external legs are present and have the same defining curl-structure!

It is interesting to stress here that we can prove no $G$ type solution can be constructed for $n>3$ and that, although we were searching in general for on-shell gauge invariant structures, all the above solutions are
gauge invariant without relying on the on-shell condition in the same way as the $H_{ij}$ structure in \eqref{H}.

This observation implies that $C_h^{\sst (n)}$ does not induce any non-trivial deformation of the gauge transformations and that the only deformations of gauge transformations are induced by 3pt structure glued together with homogeneous solutions in constructing more general particular solutions, more details on this construction will be given in \S\ref{sec:partSol}.

Eventually, the seed of the only non-trivial deformations of gauge transformations arises from the particular solution $C_p^{\sst (n)}$ that in turn can be recursively related to the cubic couplings glued together with other homogeneous solutions.
 
The fact that the only homogeneous solutions which are associated with non-trivial deformations of the gauge transformations are cubic couplings can also be rephrased in the language of differential equations. Indeed the $G$-solutions discussed around \eqref{Gstr} appear only when differential equations which we derive from the gauge invariance condition admit some singular points. 
\begin{figure}[h]
\centering
\begin{tikzpicture}
\draw [teal,fill=teal!15,dashed,very thick] (0.5,-0.6)  ellipse (2.6cm and 1.5cm);
\draw [fill=white,thick] (0.5,0.1) ellipse (1.4cm and 0.6cm);
\node at (0.5,0.1) {$H$-couplings};
\node at (0.5,-0.8) {\color{blue} \bf when $\bm\exists$ singular points:};
\node at (0.5,-1.2) {\color{red} \bf only cubic interactions};
\end{tikzpicture}
\caption{Schematic diagram for }
\label{diagram for HO}
\end{figure}
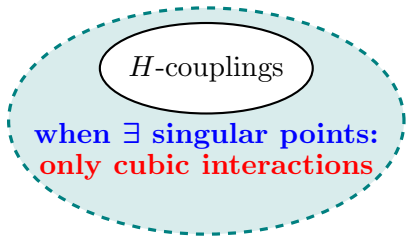
At cubic order such singular points exist in the case of three-massless-field interactions and the interactions among one massless and two massive fields with special mass differences \cite{Joung:2012rv,Joung:2012hz}. 
On the other hand, for the other cases of three-field interactions and also higher-order interactions, this never happens since the field masses (cubic case)
or Mandelstam variables prevent the coefficients of the PDE to vanish identically.

\subsection{Organisation of paper}

The paper is organised as follows. In \S\ref{sec: HO} we briefly review the structure of Noether procedure equations. In \S\ref{sec:Flat} we consider in some detail the case of Flat space interactions discussing both $n=3$ and $n=4$ explicitly. In \S\ref{DDIs} in particular we argue that for $n>3$ DDIs do not change the solution space. In \S\ref{sec:AdS} we move the analysis to constant curvature background presenting the corresponding solution. We conclude in \S\ref{sec:partSol} with a discussion of the role of the particular solution as for regards deformations of gauge transformations and with further concluding remarks.
We relegate some technical details and the examples of YM and Gravity to the Appendices \ref{app: YM}.

\bigskip

We became aware that a flat space analysis with potential overlap with the one we review in this work has been done by Stefan Fredenhagen, Olaf Kr\"uger and Karapet Mkrtchyan. We have arranged to synchronise the submission of our papers.

\section{Higher-order deformations \& Noether procedure}\label{sec: HO}

To begin with, let us consider the $n$-th order part of the gauge invariance condition of the sought interacting action:
\be
	\delta^{\sst (0)}S^{\sst (n)}
	+\delta^{\sst (n-2)}S^{\sst (2)}
	=-\left(\delta^{\sst (1)}S^{\sst (n-1)}+\,\cdots\,
	+\delta^{\sst (n-3)} S^{\sst (3)}\right),
\ee
where we assume the interaction terms $S^{\sst (2)}\,,\ldots, S^{\sst (n-1)}$  to be already determined. In order to solve the above equation, we again consider first the on-shell version:
\be
	\delta^{\sst (0)}S^{\sst (n)}
	\approx -\left(\delta^{\sst (1)}S^{\sst (n-1)}+\,\cdots\,
	+\delta^{\sst (n-3)} S^{\sst (3)}\right),
\ee
whose general solution $S^{\sst (n)}$ is the sum of the \emph{particular} solution  $S^{\sst (n)}_{p}$ and \emph{homogeneous} solution $S^{\sst (n)}_{h}$\,: 
\be
	S^{\sst (n)}=S^{\sst (n)}_{p}+S^{\sst (n)}_{h}\,,\qquad
	\left\{
	\begin{array}{c}
	\delta^{\sst (0)}S^{\sst (n)}_{p}
	\approx-\left(\delta^{\sst (1)}S^{\sst (n-1)}+\,\cdots\,
	+\delta^{\sst (n-3)} S^{\sst (3)}\right) \\
	\delta^{\sst (0)}S^{\sst (n)}_{h} \approx 0
	\end{array}\right..
\ee
Note that the particular solution is defined starting from $n=4$ and so on\,. In the following, we will find the solutions of the above equations and study their implications in an analogous manner to the cubic-interaction analysis, extending the analysis of \cite{Taronna:2011kt} to (A)dS and the analysis of \cite{Joung:2011ww} to higher-order couplings.

It is interesting to note on a side that the homogeneous solution $S_h^{\sst (n)}$ also plays the role of first non-trivial coupling in a theory with no lower-point interactions:
\begin{equation}
    S=S^{\sst (2)}+S^{\sst (n)}+S^{\sst (2n-2)}+\ldots\,.
\end{equation}
We will see how our analysis will show that cubic interaction have a special role in determining non-trivial deformations of the theory and that theories with no cubic couplings would always give rise to trivial deformations of the gauge symmetries of the free theory.\footnote{It is important to note that constructing Abelian theories of HS fields in both AdS and flat space is not a difficult task. What makes the HS problem non-trivial is the requirement of minimal coupling to gravity which requires a very special set of 3pt couplings to be turned on.}

\section{Homogeneous Solution in Flat-space: A review}\label{sec:Flat}

In this Section we review some flat space results about the Noether procedure for higher-spin fields beyond the cubic-order and in general dimension \cite{Taronna:2011kt,Taronna:2012gb,Taronna:2017wbx}. Because derivatives commute in flat space it is easier to formulate the Noether procedure problem and this will allow a more clear understanding of the extension of the flat-space result to AdS-space.

First of all it is convenient to work with the generating function formalism:
\begin{equation}
    \phi(x,u)=\frac1{\ell!}\,\phi_{\mu_1\ldots\mu_\ell}(x)u^{\mu_1}\cdots u^{\mu_\ell}\,.
\end{equation}
One can then express a generic homogeneous solution as a function of certain basic tensor contractions which are in one to one correspondence with the possible contractions of indices of the tensors $\phi_{\mu_1\ldots\mu_\ell}(x)$. Assuming parity invariance and working in generic dimension the most general TT part for a given $n$-point homogeneous solution in flat space can be therefore written as
\begin{equation}
    S_h^{(n)}=\int_{\mathbb{R}^d} C_h^{(n)}(w^{ij},y_{i}^j,z_{ij})\,\prod_{i=1}^ n\phi(x_i,u_i)\,,
\end{equation}
where we have defined the basic contractions
\begin{align}\label{vars}
    w^{ij}&=\partial_{x_i}\cdot\partial_{x_j}\,,&  y^{j}_i&=\partial_{u_i}\cdot\partial_{x_j}\,,&
    z_{ij}&=\partial_{u_i}\cdot\partial_{u_j}\,.
\end{align}
with $i\neq j$ since $i=j$ terms are removed by the TT conditions or are proportional to the mass-shell condition. Counting the number of structures one easily gets
\be
	\big(\,\#\ {\rm of}\ w\,\big)=\tfrac12\,n(n-3)\,,\quad
	\big(\,\#\ {\rm of}\ y\,\big)=n(n-2)\,,\quad
	\big(\,\#\ {\rm of}\ z\,\big)=\tfrac12\,n(n-1)\,.
	\label{num wyz}
\ee
Each local coupling would then be encoded into a polynomial in the above variables.
The variables $w^{ij}$, up to a Fourier transform, play the role of generalised Mandelstam invariants and become non-trivial only for $n\geq 4$.
The condition of gauge invariance can be formulated on the full space spanned by the variables \eqref{vars} and takes the form of the following linear PDE:
\begin{shaded}
\begin{equation}\label{gaugeinv}
    [\,C^{\sst (n)}(s,y,z)\,,\,u_{1}\cdot\partial_{x_{1}}\,]
	\approx \left(y^{1}_{i}\,\partial_{z_{i1}}+w^{1j}\,\partial_{y^{j}_{1}}\right)
	C^{\sst (n)}_{h}(w,y,z)\sim0\,.
\end{equation}
\end{shaded}
In the formulation above the symbol $\sim$ implies that the above equation should be solved on the space of formal polynomials subject to the following equivalence relations:
\begin{align}
    \chi^i_1(w,y,z)=\sum_j w^{ij}&\sim0\,,& \chi^i_2(w,y,z)=\sum_j y_i^j\sim0\,.\label{ToZeroRel}
\end{align}
This simply encodes the fact that gauge invariance should hold in the most general case up to integrations by part and boundary terms. On top of this we should also allow for couplings which become gauge invariant on the linear mass-shell and in our setting we also quotient by the TT conditions:
\begin{align}
    \chi^i_3(w,y,z)=w^{ii}&\sim0\,,& 
    \chi^i_4(w,y,z)=y_i^i&\sim0\,,& 
    \chi^i_5(w,y,z)=z_{ii}&\sim0\,.\label{TTrel}
\end{align}
In practise the problem can be reformulated as the following inhomogeneous differential equation:
\begin{equation}
    \left(y^{1}_{i}\,\partial_{z_{i1}}+w^{1j}\,\partial_{y^{j}_{1}}\right)
	C^{\sst (n)}_{h}(w,y,z)=\sum_{a=1}^5 \sum_{i=1}^n \chi_k^i(w,y,z)\,
	f_a^{(i)}(w,y,z)\,,
\end{equation}
for some generic polynomial functions $f_a^{(i)}$ which implement the equivalence relations \eqref{ToZeroRel} and \eqref{TTrel}.
Therefore, the general structure of the solution space to the above problem is given by the generic solution to the homogeneous equation which turns out to have a generic form for arbitrary $n$ to which we shall add all possible particular solutions available when varying the functions $f_a^{(i)}$ in the space of polynomials.

One way to address and solve the above set of equations is to pick a representative for the equivalence relation \eqref{ToZeroRel} and \eqref{TTrel}. The case $n=3$ is well known and was discussed in \cite{Manvelyan:2010jr,Sagnotti:2010at}. In the following we shall briefly review the $n=3$ case, we shall then address the $n=4$ example in detail and then review the extension of the analysis to $n\geq4$ \cite{Taronna:2011kt,Taronna:2012gb,Taronna:2017wbx}.

\subsection{$n=3$}
In the $n=3$ case it is easy to solve the relation \eqref{ToZeroRel} and $\eqref{TTrel}$ in terms of the following subset of independent variables:
\begin{align}
    \Xi=\{y_{1}^2,y_{2}^3,y_{3}^1,z_{12},z_{23},z_{31}\}\,,
\end{align}
where we note that $w_{ij}\approx0$ for all $i$ and $j$. The differential equation \eqref{gaugeinv} then becomes \cite{Joung:2011ww}
\begin{equation}
    (y_{2}^1\,\partial_{z_{12}}-y_{3}^1\,\partial_{z_{13}})\,C_h^{(3)}(y,z)=0\,.
\end{equation}
The characteristics of the above equation then satisfy
\begin{align}
    \frac{d y_{i}^j}{d\sigma}&=0\,,& \frac{d z_{12}}{d\sigma}&=y_2^1\,,&
    \frac{d z_{13}}{d\sigma}&=-y_3^1\,,& \frac{d z_{23}}{d\sigma}&=0\,,
    \label{3pt Gauge}
\end{align}
and the solutions can be obtained by integration as
\begin{align}
    z_{12}&=y_{2}^1\, \sigma+c_{12}\,,& z_{13}&=-y_{3}^1\, \sigma+c_{13}\,.
    \label{z 3pt}
\end{align}
Since $\frac{dC}{d\sigma}=0$,
the solution of \eqref{3pt Gauge}
is a function of $\sigma$-constants.
Besides the other trivially constant variables in $\sigma$,
one recovers an additional $\sigma$-constant,
\begin{align}\label{Gstr0}
    g&=z_{12}\,y_3^1+z_{13}\,y_{2}^1\,,
\end{align}
by eliminating the $\sigma$ dependence from
\eqref{z 3pt}.
In the end, we obtained the general 3pt solution with one massless field as \cite{Joung:2012rv} 
\begin{equation}
    C_h^{\sst (3)}=\mathcal{K}^{\sst (3)}(g,z_{23},y_{i}^j)\,.
\end{equation}

\subsection{$n=4$}
The $n=4$ example differs from the $n=3$ example from the fact that it is possible to define non-trivial Mandelstam invariants. For this reason this case will show the main features which will hold also for higher points and provides us with the main intuition which we shall generalise in the following.

The first step is to solve the equivalence relation in terms of a set of independent variables:
\begin{equation}
    \Xi=\{w^{12},\, w^{14},\,  y_{1}^2,\, y_{1}^4,\, z_{12},\,z_{13},\,z_{14},\, {\textrm{(cyclic permutations)}} \}\,,
\end{equation}
among which the variables relevant to the gauge variation of the point-splitted field $\phi_1$ 
are 
\begin{equation}
    y_{1}^2\,,\quad y_{1}^4\,\quad
    z_{12}\,,\quad z_{13}\,,\quad z_{14}\,,
    \label{rel var}
\end{equation}
whereas the rest of variables can be considered as constants. 

The gauge invariance equation then reads
\begin{align}\label{4ptGauge}
    \left(w^{12}\,\pl_{y_{1}^2}+w^{14}\,\pl_{y_1^4}+y_{2}^1\,\pl_{z_{12}}-(y_3^2+y_3^4)\,\pl_{z_{13}}+y_{4}^1\,\pl_{z_{14}}\right) C^{(4)}(w,y,z)=0\,.
\end{align}
Note that the variables 
$y_2^1,\, y_3^4, \,y_3^2,\, 
    y_4^1,\,w^{12}$ and $w^{14}$, besides the ones in \eqref{rel var}, do appear in the equation, but they can still be considered as constants
since there is no differentiation with respect to them.
The characteristics of \eqref{4ptGauge} satisfy
\begin{align}
    \frac{dy_1^2}{d\sigma}&=w^{12}\,,&  \frac{dy_1^4}{d\sigma}&=w^{14}\,,\\
    \frac{dz_{12}}{d\sigma}&=y_2^1\,,&  \frac{dz_{14}}{d\sigma}&=y_4^1\,,& \frac{dz_{13}}{d\sigma}&=-(y_3^2+y_{3}^4)\equiv y_3^1(\Xi)\,,
    \label{ODEs}
\end{align}
while all the other variables are constant in $\sigma$. The above system has 
an analogous structure as for 3pt: 
the derivatives of the $z$ variables are expressed in terms of linear combinations of $y_i^1$ and the derivatives of $y_1^i$ are expressed in terms of linear combinations of $w^{1i}$ variables. 
Since the right hand sides of the equations, namely the inhomogeneous part of the ODEs 
are constants, the solutions to \eqref{ODEs} are 
linear functions of $\sigma$.

In conclusions, since all characteristics 
in \eqref{rel var} are
linear functions of $\sigma$, we 
can make a number of constant functions
by linearly combining two of the characteristics in all possible ways and eliminating the $\sigma$ dependence. 
First, by taking linear combination of two $z_{1i}$'s
with constant coefficients $y^1_j$, we get
\begin{subequations}
\begin{align}
    h_{24}&=z_{12}\,y_{4}^1-z_{14}\,y_2^1\,,\\ 
    h_{23}&=z_{12}\,y_{3}^1-z_{13}\,y_2^1\,,\\
    h_{34}&=z_{13}\,y_{4}^1-z_{14}\,y_3^1\,. 
\end{align}
\label{h_ij}
\end{subequations}
Note in $h_{34}$ we have left implicit the dependence of $y_3^1$ on the actual independent set of variables $\Xi$ for the sake of simplicity. 
Second, by linearly combining one  $y_1^i$ 
and one $z_{1j}$
with coefficients $y^1_j$ and $w^{1i}$, we get
\begin{subequations}
\begin{align}
    h^{2}_{2}&=z_{12}\,w^{12}-y_{1}^2\,y_2^1\,,\\
    h^{4}_{2}&=z_{12}\,w^{14}-y_{1}^4\,y_2^1\,,\\
    h^{2}_{4}&=z_{14}\,w^{12}-y_{1}^2\,y_4^1\,,\\
    h^{4}_{4}&=z_{14}\,w^{14}-y_{1}^4\,y_4^1\,,\\
    h^{2}_{3}&=z_{13}\,w^{12}-y_{1}^2\,y_3^1\,,\\
    h^{4}_{3}&=z_{13}\,w^{14}-y_{1}^4\,y_3^1\,.
\end{align}
\label{h^i_j}
\end{subequations}
Finally, by taking
linear combination of two $y_1^i$'s
with coefficients $w^{1j}$, we get
\begin{equation}
   h^{24}=y_{1}^2\,w^{14}-y_1^4\,w^{12}\,.
   \label{h^ij}
\end{equation}
The general solution of \eqref{4ptGauge} is then given by any polynomial function
\begin{align}
    C^{\sst (4)}=\mathcal{K}^{\sst (4)}(h_{ij},h^j_i,h^{ij},c_a)\,,
\end{align}
where $c_a$ are all the constant characteristics
and 
$h_{ij}, h^i_j$ and $h^{ij}$ are 
\begin{subequations}
\begin{align}
    h_{ij}&=z_{1i}\,y_{j}^1-z_{1j}\,y_i^1\,,\\ 
    h^j_i&=z_{1i}\,w^{1j}-y_{1}^j\,y_i^1\,,\\
    h^{ij}&=y_{1}^i\,w^{1j}-y_1^j\,w^{1i}\,.
\end{align}
\label{h structure}
\end{subequations}
Note that comparing the above to the lists in
\eqref{h_ij}, \eqref{h^i_j}
and \eqref{h^ij},
we find additional structures
but they are in fact related to the others by
\begin{equation}
    h^3_2=-h^2_2-h^4_2\,,\quad
    h^3_3=-h^2_3-h^4_3\,,\quad
    h^3_4=-h^2_4-h^4_4\,,\quad
    h^{32}=-h^{34}=h^{24}\,.
\end{equation}
Therefore, we see that all non-trivial solutions of \eqref{4ptGauge}
are given by the $h$-structures \eqref{h structure}
up to the above redundancies.

A general consequence of the above finding 
is that all gauge invariant 4pt interactions
do not induce a deformation of gauge symmetry.
To see this point, it is enough to realise that 
the $h$-structures \eqref{h structure} 
are all based on curl-type operations:
\begin{subequations}
\begin{align}
    h_{ij}&=2\,(\partial_{u_i})_{\mu}\,(\partial_{u_j})_{\nu}\,
    (\partial_{u_1})^{[\m}\,(\partial_{x_1})^{\n]},\\ 
    h^j_i&=2\,(\partial_{u_i})_\m\,(\partial_{x_j})_{\n}\,
    (\partial_{u_1})^{[\m}\,(\partial_{x_1})^{\n]},\\
    h^{ij}&=2\,(\partial_{x_i})_\m\,(\partial_{x_j})_{\n}\,
    (\partial_{u_1})^{[\m}\,(\partial_{x_1})^{\n]}\,,
\end{align}
\label{h curl}
\end{subequations}
where $T^{[\m\n]}=\frac12(T^{\m\n}-T^{\n\m})$\,.
This makes gauge invariance manifest even without imposing the equivalence relations \eqref{ToZeroRel} and \eqref{TTrel} simply because
\begin{equation}
        \big[\,(\partial_{u_1})^{[\m}\,(\partial_{x_1})^{\n]}\,,\,
        u_1\cdot\partial_{x_1}\,\big]=0\,.
\end{equation}
It is interesting to also compare the $n=4$ result with the result obtained at 3pt. The basic structure of the solution is very similar and the $h$-structures we obtained at $n=4$ play the role of straightforward generalisations of the $g$ structure \eqref{Gstr0}. However the main difference is that while the $y_1^2$ structures was associated with a constant characteristic in $\sigma$ for $n=3$, the $y_1^j$ structures are not anymore constant for $n=4$ but they can only appear in the combination given by the $h$-structures \eqref{h^i_j}. This implies a key difference between $n=3$ and $n=4$: \emph{All solutions to the gauge invariance conditions are gauge invariant off-shell, without using neither the TT or the mass-shell condition.} We therefore recover as a corollary that no non-trivial deformation of the gauge symmetries of the free theory can be induced by a quartic coupling which solves the homogeneous equation $\delta^{(0)}S_h^{(n)}\approx0$.\footnote{See \cite{Taronna:2017wbx} for previous works where similar statements were made.}

Let us reiterate the difference between 3pt and 4pt couplings
with the example
of the interactions involving only one massless field.
The PDE for the cubic interaction reads
\be
	\left[y^{1}_{2}\,\partial_{z_{21}}-y^{2}_{3}\,\partial_{z_{31}}+
	\left( m_{2}^{\,2}-m_{3}^{\,2} \right)\partial_{z^{3}_{1}}
	\right] C^{\sst (3)}=0\,,
\ee
displaying a singular point at $m_{2}^{\,2}-m_{3}^{\,2}=0$. On the other hand the PDE for quartic coupling is given by the following expression
\be	
	\left[y^{1}_{2}\,\partial_{z_{21}}
	+y^{1}_{3}\,\partial_{z_{31}}-(y^{2}_{4}+y^{3}_{4})\,\partial_{z_{41}}
	-t\,\partial_{y_{1}^{3}}-
	\tfrac12(s-m_{2}^{\,2}-m_{3}^{\,2}-m_{4}^{\,2})\,
	\partial_{y^{4}_{1}}
	\right] C^{\sst (4)}_{h}=0\,.
\ee
Here, the appearance of the Mandelstam variables for the first time at $n=4$ allows to avoid any singular point, the latter becoming of measure zero. One can summarise the above facts considering the generic form of the gauge-consistency PDE:
\be
	\left( y_{a}\,\mathbb A^{ab}\,\partial_{z_{b}}
	-\mathbb B^{c}\,\partial_{y_{c}} \right)C=0\,.
\ee
where depending on the structure of the coefficient $\mathbb B$ one has two cases:
\begin{itemize}
\item
$\mathbb B=0$ : Singular points of DE 
	\quad $\Rightarrow$ \quad 
	Both \textbf{on}/\textbf{off}-shell invariant couplings
\item
$\mathbb B\neq 0$ :  No singular point 
	\quad $\Rightarrow$ \quad
	Only \textbf{off}-shell gauge invariant couplings
\end{itemize}
All in all, the presence of Mandelstam variables within the coefficient $\mathbb B$ starting at $n=4$ implies that the PDE does not have any singular point. Hence, this excludes the possibility of having further couplings with respect to the ones described above and proves also the non-deforming nature of the homogeneous solutions for $n=4$. The only homogeneous solutions that can be non-Abelian arise at the cubic level due to the \textit{non-trivial} singular points of the corresponding gauge-invariance PDE.

In the following we shall review the generalisation of these statements to arbitrary $n\geq 4$.

\subsection{General $n$}

The discussion for arbitrary $n$ follows closely the $n=4$ case. Indeed the main observation is that after solving the equivalence relations \eqref{ToZeroRel} and \eqref{TTrel} the characteristics of the gauge invariance condition have the same structure as for $n=3,4$:
\begin{align}
    \frac{dy_{1}^i}{d\sigma}&=w^{1i}\,,&  \frac{dz_{1}^i}{d\sigma}&=y_i^1\,,
\end{align}
where $i\neq1$. Notice that the above structure of the characteristics is not affected by the choice of independent variables $\Xi$,
though upon a particular choice some of $y_i^1$ 
become linear combinations of the rest.
Instead, $y_i^1$ 
can never be related to $y_1^i$ by the equivalence relations \eqref{ToZeroRel} and \eqref{TTrel}. Therefore, 
it is safe to work with the linearly dependant variables without specifying a basis. 

The general solution of the gauge invariance conditions then reads for arbitrary $n$ as
\begin{equation}
    C^{(n)}=\cK^{(n)}(h_{ij},h^j_i,h^{ij},c_a)\,,
    \label{n pt}
\end{equation}
where $c_a$ are all constant characteristics and where $h_{ij}$,
$h_i^j$ and $h^{ij}$ are again defined as \eqref{h structure}
but now $i,j$ runs from 2 to $n$.
As in the $n=4$ case, not all $h$-structures are independent
but they are subject to linear relations,
\begin{equation}
    \sum_{i=2}^n h^{ki}=0\,,
    \qquad 
    \sum_{i=2}^n h_k^i=0\,.
    \label{h rel}
\end{equation}
In the end, we conclude that the result for $n=4$ straightforwardly generalises to any $n$. At this point a few comments are in order:
\begin{itemize}
    \item All $n$-pt couplings 
    gauge invariant with respect to one massless higher-spin field, $\phi(x_1,u_1)$ here, are gauge invariant off-shell (again due to the fact that $y_1^i$ structure is not associated to a constant characteristic) for $n\ge 4$.
    \item Since such gauge invariant $n$-pt couplings are polynomials of $h$-structures \eqref{h structure} which are based on curls \eqref{h curl}, the coupling function $\cK^{(n)}$ involves $\partial_{u_1}$ only through the curl
    $[\partial_{x_1}\,\partial_{u_1}]^{\m\n}$.
    In other words, the components of the massless higher-spin field
    are all saturated by curls.
    From this, we can conclude that
    the massless higher-spin field enters to such couplings
    only through the form of linearised curvature.
    \item So far we assumed that all the fields entering to the interaction are massless, and this assumption was used in \eqref{TTrel}.
    When some of the fields become massive, 
    only the relation \eqref{TTrel} gets modified
    to the massive mass-shell condition.
    However this relation is never used in the derivation of the 
    solution, so our solution \eqref{n pt}
    is valid irrespectively of the mass of the other fields but 
    the massless field $\phi(x_1,u_1)$.
    Remark that this was not the case for the 3pt coupling of a massless field to 
    two massive fields with equal mass 
    where the $y_1^i$ characteristics collapses to a constant
    so can appear as an independent solution.
    
\end{itemize}

\subsection{Comments on Dimensional-dependent identities}\label{DDIs}

So far in our analysis we have been working in arbitrary dimensions and have not considered dimensional dependent identities (DDIs) 
generated by over-antisymmetrisations of a number of indices greater than the space-time dimension. 
For the homogeneous $n$-pt coupling of totally symmetric fields,
DDIs can be generated as
\begin{align}
    \mathsf{A}^{[i_1\cdots i_p][j_{p+1}\cdots j_{d+1}]}_{[k_1\cdots k_q][l_{q+1}\cdots l_{d+1}]}=&\left[(\partial_{u_{i_1}})_{[\mu_{1}}\cdots (\partial_{u_{i_p}})_{\mu_{p}}
    (\partial_{x_{j_{p+1}}})_{\mu_{p+1}}\cdots (\partial_{x_{j_{d+1}}})_{\mu_{d+1]}}\right]\times
    \nonumber\\
    &\times\left[(\partial_{u_{k_1}})^{[\mu_1}\cdots (\partial_{u_{k_{q}}})^{\mu_q}
    (\partial_{x_{l_{q+1}}})^{\mu_{q+1}}\cdots (\partial_{x_{l_{d+1}}})^{\mu_{d+1}]}\right]\equiv 0\,,
    \label{DDI}
\end{align}
where $i_\bullet, j_\bullet, k_\bullet, l_\bullet$ take values in $\{1, \ldots n\}$.
Making use of the product identities of Levi-Civita symbols,
DDIs can be expressed as degree $d+1$ polynomials of $z_{ij}$, $y_i^j$ and $w^{ij}$\,,
and 
these polynomials are identically zero.
For the existence of DDIs, as is clear from the expression \eqref{DDI},
the number of $\partial_{u_i}$ and $\partial_{x_i}$ should 
not be smaller
than $d+1$.
Taking into account the total derivative, this implies
 that non-trivial dimensional DDIs arise only
for the $n$-point interactions with
\begin{equation}
    n\ge \frac{d+2}2\,.
    \label{DDI n}
\end{equation}
This bound is in agreement with the existence of DDIs for 3pt couplings in $d\leq 4$ and with the existence of DDIs for 4pt couplings in $d\leq 6$. A similar counting gives a bound for couplings proportional to the $\epsilon$-tensor. They can exist only if $2n-1\geq d$ implying that the highest dimension in which a parity violating quartic interaction exist is $d=7$. Similarly the highest dimension for a parity violating cubic interaction is $d=5$. In any higher-dimension the epsilon tensor cannot be used to contract non-trivial quartic or cubic interactions respectively.\footnote{The fact that $d=7$ is the highest dimension in which $\epsilon$ tensor contributes to 4pt structures may have non-trivial implications for the bootstrap program and is consistent with the fact that bootstrap equations for CFT correlators involving symmetric tensors become redundant in $d\leq 6$. This discontinuity in the number of independent bootstrap equations arising at the level of stress tensor correlators/totally symmetric fields might be one of the reason why it is so hard to construct non-trivial interacting unitary CFTs above $d=6$.}

It is clear from its definition \eqref{DDI} that the gauge variation a DDI is a DDI, so the gauge variation of a coupling involving a DDI also involves a DDI: the gauge variation of an identically vanishing coupling vanishes identically.  
Below we will argue that if $n\ge 4$ also the opposite holds: if the gauge variation of a certain coupling is proportional to a DDI, then the coupling itself must be proportional to a DDI.

In order to understand the situation more clearly,
let us begin by removing the ambiguities of field redefinitions and  total derivaties in 
the TT part of the $n$-point coupling $C^{\sst (n)}(w,y,z)$.
Having in mind to analyse the gauge invariance 
for the first field, we can restrict $C^{\sst (n)}$ to 
depend only on the following set of variables,
\ba
    A=&&\big\{w^{ij} \,\big|\, 1\le i<j\le n-2\big\}
    \cup \big\{w^{i(n-1)}\,\big|\, 1\le i\le n-3\big\} \nn
    &&\cup \,\big\{y_i^j\,\big|\, 1\le i\neq j\le n-1\big\} 
    \cup \,\big\{y_n^j\,\big|\, 1\le j\le n-2\big\}
    \nn
    &&
    \cup \big\{z_{ij}\,\big|\, 1\le i<j\le n\big\}\,.
    \label{var set}
\ea
Then the gauge variation with respect to the first field is
\be 
    \left(\delta^{\sst (0)}C^{\sst (n)}\right)(w,y,z)= \left( \sum_{i=2}^n y_i^1\,\pl_{z_{1i}}+
    \sum_{j=2}^{n-1} w^{1j}\,\pl_{y_1^j}\right)
    C^{\sst (n)}(w,y,z).
\ee
Note that the gauge variation introduces $y_i^1$ 
with $2\le i \le n$ and $w^{1j}$ with $2\le j\le n-1$.
The former  is already in the set $A$ \eqref{var set},
whereas the latter is so only when $n\ge 4$\,.
Hence, one can now distinguish two cases: 
(i) $w^{1j}\neq 0$ which happens for $n\geq4$, and
(ii) $w^{1j}\approx 0$ which happens for $n=3$. 

\paragraph{Case $n\ge 4$}

In this case, we just need to solve the equation,
\be 
   \left( \sum_{i=2}^n y_i^1\,\pl_{z_{1i}}+
    \sum_{j=2}^{n-1} w^{1j}\,\pl_{y_1^j}\right)
    C^{\sst (n)}(w,y,z)=0\,.
    \label{ginv}
\ee
The function $C^{\sst (n)}(w,y,z)$ can
be expressed as
\begin{align}
    C(w,y,z)
    =C_{\mu_1^{1}\cdots\mu_{r_1}^{1};\nu_1^{1}\cdots\nu_{s_1}^{1};\cdots}\left(\partial_{x_1}^{\mu_1^{1}}\cdots \partial_{x_1}^{\mu^1_{r_1}}\right)\left(\partial_{u_1}^{\nu_1^{1}}\cdots \partial_{u_1}^{\nu^1_{s_1}}\right)\cdots\,,
\end{align}
in terms of the tensor $C_{\mu_1^{1}\cdots\mu_{r_1}^{1};\nu_1^{1}\cdots\nu_{s_1}^{1};\cdots}$.
 The differential equation \eqref{ginv}  
is equivalent to 
the Young symmetrization condition of the tensor,
\be
    C_{(\mu_1^{1}\cdots\mu_{r_1}^{1};\nu_1^{1})\nu_2^{1}\cdots\nu_{s_1}^{1};\cdots}=0\,.
    \label{C Young sym}
\ee 
Let us look into this point more closely.
The coupling tensor $C_{\mu_1^{1}\cdots\mu_{r_1}^{1};\nu_1^{1}\cdots\nu_{s_1}^{1};\cdots}$ is symmetic in each of $\mu^i_1\cdots \mu^i_{r_i}$ and 
$\nu^j_1\cdots \nu^j_{s_j}$, hence 
it has the symmetry of the Young tableaux, 
\ba 
    && C_{\mu_1^{1}\cdots\mu_{r_1}^{1};\nu_1^{1}\cdots\nu_{s_1}^{1};
   \mu_1^{2}\cdots\mu_{r_2}^{2}; \,\cdots ;\nu_1^{n}\cdots\nu_{s_n}^{n}}\nn 
    &&
    \sim 
    \begin{ytableau}
    \mu_1^1 & \cdots  & \mu_{r_1}^1
    \end{ytableau}
    \otimes 
     \begin{ytableau}
    \nu_1^1 & \cdots  & \nu_{s_1}^1
    \end{ytableau}
    \otimes 
    \begin{ytableau}
    \mu_1^2 & \cdots & \mu_{r_2}^2
    \end{ytableau}
     \otimes 
    \cdots
    \otimes 
     \begin{ytableau}
    \nu_1^n & \cdots  & \nu_{s_n}^n
    \end{ytableau}\,.
    \label{C tab}
\ea 
The above tensor product can be decomposed into irreducible representations
by successively applying the Littlewood-Richardson  rule.
In the end, we will obtain
the direct sum,
\be 
    \bigoplus_{\mathbb Y\in Y} \mathbb Y\,,
\ee 
where $Y$ is the set of the 
Young tableaux appearing in the decomposition of \eqref{C tab}.
Eventually, each $\mathbb Y$
corresponds to a possible interaction vertex,
but we should impose a few more restrictions.

\begin{itemize}

\item The coupling tensor $C_{\mu_1^{1}\cdots\mu_{s_1}^{1};\nu_1^{1}\cdots\nu_{p_1}^{1};\ldots}$ 
is made by the metric tensors $\eta_{\nu^i\nu^j}, \eta_{\mu^i\nu^j}$ and $\eta_{\mu^i\mu^j}$
due to Lorentz invariance.
This was already reflected in the choice of the variables $w, y, z$,
\be 
    w^{ij}=\eta_{\nu^i\nu^j}\,\partial_{x_i}^{\nu^i}\,
    \partial_{x_j}^{\nu^j}\,,
    \qquad
    y_{i}^{j}=\eta_{\mu^i\nu^j}\,\partial_{u_i}^{\mu^i}\,
    \partial_{x_j}^{\nu^j}\,,
    \qquad 
    z_{ij}=\eta_{\mu^i\mu^j}\,\partial_{u_i}^{\mu^i}\,
    \partial_{u_j}^{\mu^j}\,.
\ee 
Therefore, not all Young tableaux 
$\mathbb Y \in Y$ 
suit the coupling tensor, but the ones that
can be found in the tensor product 
decompositions of the metric tensors.
It is known that the Young tableaux appearing in this decomposition have rows with an even number of boxes:
let us call the set of such Young tableaux
as $Y_{\rm even}$.
For any Young tableau $\mathbb Y \in Y_{\rm even}$,
there is a unique (up to an overall constant) tensor $\eta_{\,\mathbb Y}$ of the 
corresponding symmetry 
made by metric tensors, 
and the coupling tensor 
can be expressed as a linear combination
of $\eta_{\,\mathbb Y}$:
\be 
    C_{\mu_1^{1}\cdots\mu_{r_1}^{1};\nu_1^{1}\cdots\nu_{s_1}^{1};\cdots}
    =\sum_{\mathbb Y\in Y_{\rm even}}
    \alpha_{\mathbb Y}\,\eta_{\,\mathbb Y}\,.
\ee
Here $\alpha_{\mathbb Y}$ are coefficients parameterising the linear combination of couplings.

\item The function $C(w,y,z)$ is restricted to the variables $A$ \eqref{var set}. This means that the tensor $C_{\mu_1^{1}\cdots\mu_{r_1}^{1};\nu_1^{1}\cdots}$
does not involve $\eta_{\bullet\bullet}$ corresponding to the omitted variables.
Without loss of generality, this will reduce possible form of the coupling tensor to
\be 
    C_{\mu_1^{1}\cdots\mu_{r_1}^{1};\nu_1^{1}\cdots}
    =\sum_{m} c_m\,
    B^{m}_{\mu_1^{1}\cdots\mu_{r_1}^{1};\nu_1^{1}\cdots}
\ee    
where 
\be 
    B^{m}_{\mu_1^{1}\cdots\mu_{r_1}^{1};\nu_1^{1}\cdots}=
    \sum_{\mathbb Y\in Y_{\rm even}}
    \beta^{m}_{\,\mathbb Y}\,\eta_{\,\mathbb Y}\,,
\ee
are the most general linear combinations
of $\eta_{\mathbb Y}$
which do not involve  $\eta_{\bullet\bullet}$ corresponding to the omitted variables.
Hence $B^{m}_{\mu_1^1\cdots \mu_{r_1}^1;\nu_1^1\cdots}$ 
form the basis of the coupling tensor
(but we did not impose yet the gauge invariance condition).
A very simple way to understand the basis tensors $B^{m}_{\mu_1^1\cdots \mu_{r_1}^1;\nu_1^1\cdots}$
is viewing them as the monomial basis of the variables in $A$
so that they span the space of function $C$.

\item  To have a concrete idea on the above points, let us consider
a toy example of $C_{\mu_1\mu_2;\nu_1\nu_2}$:
\ba 
    C_{\mu_1\mu_2;\nu_1\nu_2}
    && \sim 
    \begin{ytableau}
    \mu_1 & \mu_2
    \end{ytableau}\, \otimes 
    \,  \begin{ytableau}
    \nu_1 & \nu_2
    \end{ytableau} \nn 
    &&
    =\,
     \begin{ytableau}
    \mu_1 & \mu_2 & \nu_1 & \nu_2
    \end{ytableau}\,\oplus\,
     \begin{ytableau}
    \mu_1 & \mu_2 & \nu_1 \\ \nu_2
    \end{ytableau}\,\oplus\,
     \begin{ytableau}
    \mu_1 & \mu_2 & \nu_2 \\ \nu_1
    \end{ytableau} \,\oplus\,
    \begin{ytableau}
    \mu_1 & \mu_2 \\ \nu_1 & \nu_2
    \end{ytableau}\,.
\ea
Among the above, the Young tableaux in $Y_{\rm even}$
are
\be 
    \begin{ytableau}
    \mu_1 & \mu_2 & \nu_1 & \nu_2
    \end{ytableau}\,\oplus\,
    \begin{ytableau}
    \mu_1 & \mu_2 \\ \nu_1 & \nu_2
    \end{ytableau}\,,
\ee 
and the corresponding $\eta_{\,\mathbb Y}$ tensors 
are
\ba 
    \eta_{\,\tiny \begin{ytableau}
    \mu_1 & \mu_2 & \nu_1 & \nu_2
    \end{ytableau}}
    \eq \frac13\left(\eta_{\mu_1\mu_2}\,\eta_{\nu_1\nu_2}
    +\eta_{\mu_1\nu_2}\,\eta_{\nu_1\mu_2}+
    \eta_{\mu_1\nu_1}\,\eta_{\mu_2\nu_2}\right), \\
     \eta_{\,\tiny \begin{ytableau}
    \mu_1 & \mu_2 \\ \nu_1 & \nu_2
    \end{ytableau}}
    \eq \frac13\left(2\,\eta_{\mu_1\mu_2}\,\eta_{\nu_1\nu_2}
    -\eta_{\mu_1\nu_2}\,\eta_{\nu_1\mu_2}-
    \eta_{\mu_1\nu_1}\,\eta_{\mu_2\nu_2}\right).
\ea
Therefore, we first find
\be 
    C_{\mu_1\mu_2;\nu_1\nu_2}=
    \alpha_1\, \eta_{\,\tiny \begin{ytableau}
    \mu_1 & \mu_2 & \nu_1 & \nu_2
    \end{ytableau}}
    +\alpha_2\,\eta_{\,\tiny \begin{ytableau}
    \mu_1 & \mu_2 \\ \nu_1 & \nu_2
    \end{ytableau}}\,.
\ee 
Impose the condition that
the tensor $C_{\mu_1\mu_2;\nu_1\nu_2}$
does not involve, say $\eta_{\mu_1\mu_2}$,
the coupling tensor is reduced to
\be
    C_{\mu_1\mu_2;\nu_1\nu_2}=
    c_1\,B^1_{\mu_1\mu_2;\nu_1\nu_2}\,,
\ee 
\be     
    B^1_{\mu_1\mu_2;\nu_1\nu_2}
    =2\,
      \eta_{\,\tiny \begin{ytableau}
    \mu_1 & \mu_2 & \nu_1 & \nu_2
    \end{ytableau}}
    -\eta_{\,\tiny \begin{ytableau}
    \mu_1 & \mu_2 \\ \nu_1 & \nu_2
    \end{ytableau}}
    =\eta_{\mu_1\nu_2}\,\eta_{\nu_1\mu_2}+
    \eta_{\mu_1\nu_1}\,\eta_{\mu_2\nu_2}\,.
\ee 

\end{itemize}

Now we impose the condition \eqref{C Young sym}.
We realise the symmetrization of indices
$\mu_1^1\cdots \mu_{r_1}^1 \nu_1^1$ 
as the action of the operator $\mathsf S$ on the tensor:
\be 
    C_{(\mu_1^1\cdots \mu_{r_1}^1;\nu_1^1)\nu_2^1\cdots }
    =\mathsf S\,C_{\mu_1^1\cdots \mu_{r_1}^1;\nu_1^1\cdots }
    =\sum_m c_m\,\mathsf S\,
    B^m_{\mu_1^1\cdots \mu_{r_1}^1;\nu_1^1\cdots }\,.
\ee 
and
\be
    \mathsf S\,
    B^m_{\mu_1^1\cdots \mu_{r_1}^1;\nu_1^1\cdots }=\sum_{\mathbb Y\in Y_{\rm even}}
    \beta^m_{\,\mathbb Y}
    \,\mathsf S\,\eta_{\,\mathbb Y}\,.
\ee 
Young tableaux of the same Young diagram
form an irreducible representation under the symmetric group, hence
\be 
    \mathsf S\,\eta_{\,\mathbb Y}
    =\sum_{\mathbb Y'\in Y_{\rm even}} 
    \sigma_{\mathbb Y,\mathbb Y'}\, \eta_{\,\mathbb Y'}\,,
\ee 
where the coefficients $\sigma_{\mathbb Y,\mathbb Y'}$
vanish unless
the Young tableaux $\mathbb Y'$
and $\mathbb Y$ have the same shape, i.e. Young diagram.
The above equation determines how each basis tensor $B^m_{\mu_1^1\cdots \mu_{r_1}^1;\nu_1^1\cdots }$ transforms under $\mathsf S$.
The fact that the symmetrization $\mathsf S$
does not introduce any new variables 
means that $\mathsf S$ is an endomorphism
of the space of tensors spanned by $B^m_{\mu_1^1\cdots \mu_{r_1}^1;\nu_1^1\cdots }$\,:\footnote{If we regard the tensors $B^m_{\mu_1^1\cdots \mu_{r_1}^1;\nu_1^1\cdots }$ as
monomials of the variables in $A$, one should keep in mind 
that such a tensor corresponds to different monomials 
depending on whether it is viewed as a basis of coupling function
or a basis of gauge variation of coupling function. }
\be 
    \mathsf S\,B^m_{\mu_1^1\cdots \mu_{r_1}^1;\nu_1^1\cdots }
    =\sum_{m'}M_{m,m'}
     \,B^{m'}_{\mu_1^1\cdots \mu_{r_1}^1;\nu_1^1\cdots }\,.
     \label{S B}
\ee 
In other words, the vector space $V$ of coupling tensors 
satisfy $\mathsf S\,V\subset V$\,.
Therefore, the condition \eqref{C Young sym}
is equivalent to finding 
the kernel of the matrix $S_{m,m'}$\,:
\be 
    \sum_m c_m\,S_{m,m'}=0\,.
\ee 
At this point, let us come back to our original question:
whether there can be
any coupling $C$ whose gauge variation is a DDI but itself is not.
In the our setting, DDIs are simply,
\be 
    \eta_{\,\mathbb Y}=0\,,\qquad 
    \forall \,\mathbb Y \in Y_{\rm DDI}\,.
\ee 
where $Y_{\rm DDI}\subset Y_{\rm even}$ is
the set of Young tableaux
having more than $d$ rows.
Suppose we have a coupling tensor $C_{\mu_1^1\cdots \mu_{r_1}^1;\nu_1^1\cdots }$
satisfying
\be 
     \mathsf S\,C_{\mu_1^1\cdots \mu_{r_1}^1;\nu_1^1\cdots }
    =\sum_{\mathbb Y\in Y_{\rm DDI}} 
    d_{\mathbb Y}\,
     \eta_{\,\mathbb Y}\,.
     \label{DDI coupling}
\ee 
Since the symmetrization $\mathsf S$ is a projection, $\mathsf S^2=\mathsf S$\,,
we have 
\be
    \mathsf S\left(\sum_{\mathbb Y\in Y_{\rm DDI}} 
    d_{\mathbb Y}\,
     \eta_{\,\mathbb Y}\right)
     =\sum_{\mathbb Y\in Y_{\rm DDI}} 
    d_{\mathbb Y}\,
     \eta_{\,\mathbb Y}\,.
     \label{S ddi}
\ee 
This implies that to such a coupling tensor $C^m_{\mu_1^1\cdots \mu_{r_1}^1;\nu_1^1\cdots }$
corresponds an ordinary one (that is, a coupling gauge invariant without relying on DDI),
\be 
    \tilde C_{\mu_1^1\cdots \mu_{r_1}^1;\nu_1^1\cdots }
    =C_{\mu_1^1\cdots \mu_{r_1}^1;\nu_1^1\cdots }-\sum_{\mathbb Y\in Y_{\rm DDI}} 
    d_{\mathbb Y}\,
     \eta_{\,\mathbb Y}\,,
     \qquad 
     \mathsf S\, \tilde C_{\mu_1^1\cdots \mu_{r_1}^1;\nu_1^1\cdots }=0\,.
\ee
What is important here is
that $\sum_{\mathbb Y\in Y_{\rm DDI}}$
can be viewed as a (trivial) coupling tensor 
since it is a linear combination of $B^m_{\mu_1^1\cdots \mu_{r_1}^1;\nu_1^1\cdots }$ 
(this is due to \eqref{S B} and to the fact that $\sum_{\mathbb Y\in Y_{\rm DDI}} 
    d_{\mathbb Y}\,
     \eta_{\,\mathbb Y}$ is in the image of $\mathsf S$).

This issue can be also understood from a slightly different point of view: recall that 
the action $\mathsf S$ does not change 
the Young diagram, hence if we begin
with the basis $B^m_{\mu_1^1\cdots \mu_{r_1}^1;\nu_1^1\cdots }$
where we delete all $\eta_{\,\mathbb Y \in Y_{\rm DDI}}$,\footnote{This step
can only introduce some linear dependencies
in $B^m_{\mu_1^1\cdots \mu_{r_1}^1;\nu_1^1\cdots }$'s. Hence, we can just reduce the coefficient $c_m$ accordingly.}
then the symmetrization action $\mathsf S$
will never generate any new $\eta_{\,\mathbb Y \in Y_{\rm DDI}}$.
In either way, we can conclude that
it is not possible to have a coupling tensor $C_{\mu_1^1\cdots \mu_{r_1}^1;\nu_1^1\cdots }$
which is not a (linear combination of) DDI,
but $C_{(\mu_1^1\cdots \mu_{r_1}^1;\nu_1^1)\cdots }$ is.

Let us close the analysis of the $n\ge 4$ case
by showing how the kernel can be 
identified in a constructive manner.
In fact, the condition \eqref{C Young sym} can be solved at the stage
of applying Littlewood-Richardson rule to the first two Young tableaux:
we restrict to
\ba 
    && C_{\mu_1^{1}\ldots\mu_{r_1}^{1};\nu_1^{1}\ldots\nu_{s_1}^{1};
   \mu_1^{2}\ldots\mu_{r_2}^{2};
    \cdots ;\nu_1^{n}\ldots\nu_{s_n}^{n}
    } \nn 
    &&\sim 
    \begin{ytableau}
    \mu_1^1 & \cdots & \cdots & \cdots & \mu_{r_1}^1\\
    \nu_1^1 & \cdots  & \nu_{s_1}^1
    \end{ytableau}
    \otimes 
    \begin{ytableau}
    \mu_1^2 & \cdots & \mu_{r_2}^2
    \end{ytableau}
    \otimes 
    \cdots \otimes 
     \begin{ytableau}
    \nu_1^n & \cdots  & \nu_{s_n}^n
    \end{ytableau}.
    \label{sol C tab}
\ea
By applying further the Littlewood-Richardson rules,
 the direct sum, $\bigoplus_{\mathbb Y\in \tilde Y}\mathbb Y$
 where $\tilde Y\subset Y$ is the 
 set of Young tableaux satisfying $\mathsf S\,\mathbb Y=0$.
 Then, we further restrict to
 the set $\tilde Y_{\rm even}$ even-row Young tableaux 
 and eventually take the subspace
 which do not involve $\eta_{\bullet\bullet}$
 corresponding to the omitted variables.

\paragraph{Case $n=3$}

In this case, the arguments of $n\ge 4$ do not apply any more,
and there may exist non-trivial couplings whose
gauge variation vanishes up to DDIs.
From the condition \eqref{DDI n}, the relevant dimensions are $d\le 4$.
In $d=4$, some massless cubic interactions become proportional to DDI leaving always two independent interactions for each $s_1,s_2, s_3$ couplings,
but there is no interaction whose gauge variation is a DDI.
In $d=3$, the Chern-Simons formulation tells us
that we have two derivative 
$s-s-2$ interactions, whose analogue cannot be found in
dimensions higher than 3.
In fact, these two-derivative interactions are precisely the one
whose gauge variation is proportional to DDI.
Systematic analysis of 3d interaction vertices
has been carried out in \cite{Mkrtchyan:2017ixk,Kessel:2018ugi}
(see also \cite{Blencowe:1988gj,Campoleoni:2012hp,Fredenhagen:2018guf,Fredenhagen:2019hvb})
and all interactions whose gauge variation becomes a DDI have been explicitly classified.
Here, we revisit this story to see more concretely why 
this type of interactions exist only for the cubic interactions
but not for higher order interactions.

We begin again with the set of variable $A$ \eqref{var set}, which reduces to
\be 
    A=\{y_1^2, y_2^1, y_3^1, z_{12}, z_{13}, z_{23} \}. 
\ee 
The gauge variation of the coupling function $C$ is
\be 
    \left(\delta^{\sst (0)}C^{\sst (3)}\right)
    (w,y,z)
    =\left(y_2^1\,\partial_{z_{12}}
    +y_3^1\,\partial_{z_{13}}
    +w^{12}\,\partial_{y_1^2}\right)
    C(y,z)\,,
\ee 
and we find that the gauge variation introduces back the variable
$w^{12}$ which has been discarded from the set $A$.
Hence, this means that the vector space $V$ of coupling tensors
is not invariant under the action of $\mathsf S$\,:
$\mathsf S\,V\not\subset V$.
For the subsequent analysis, it is useful explicitly consider couplings involving $w^{12}$.
But the monomial functions involving $w^{12}$ correspond
to different coupling tensors depending on 
whether one considers a coupling or a gauge variation of a coupling. This is because additional $w^{12}$ monomials are generated by the gauge variation. Hence we have schematically
\be
    V\oplus W = \bar V\oplus \bar W\,.
\ee 
where $V$ and $\bar V$ are the vector spaces of tensors
associated with couplings and variation of couplings which do not 
involve $\eta_{\bullet\bullet}$
corresponding to $w^{12}$.
And $W$ and $\bar W$ are the analoguous vector spaces
where the tensors do involve $\eta_{\bullet\bullet}$
corresponding to $w^{12}$.
In terms of these, we have
\be 
    \mathsf S\,V\subset \bar V\oplus \bar W\,.
    \label{V W space}
\ee 
If the basis tensors of the space $V, \bar V, W$ and $\bar W$ are 
denoted by 
$B^m_{\mu_1^1\cdots \mu_{r_1}^1;\nu_1^1\cdots},
D^a_{\mu_1^1\cdots \mu_{r_1}^1;\nu_1^1\cdots},$
$\bar  B^{\bar m}_{\mu_1^1\cdots \mu_{r_1}^1;\nu_1^1\cdots}$
and $\bar  D^{\bar a}_{\mu_1^1\cdots \mu_{r_1}^1;\nu_1^1\cdots}$\,,
respectively,
then \eqref{V W space} implies
\be 
    \mathsf S\,B^m_{\mu_1^1\cdots \mu_{r_1}^1;\nu_1^1\cdots }
    =\sum_{\bar m}S_{m,\bar m}
     \,\bar B^{\bar m}_{\mu_1^1\cdots \mu_{r_1}^1;\nu_1^1\cdots }
     +\sum_{\bar a} S_{m,\bar a}
      \,\bar D^{\bar a}_{\mu_1^1\cdots \mu_{r_1}^1;\nu_1^1\cdots }
     \,.
\ee
Let us consider now a coupling gauge invariant up to DDIs:
the corresponding coupling tensor $C_{\mu_1^1\cdots \mu_{r_1}^1;\nu_1^1\cdots }$ should
satisfy
\be 
     \mathsf S\,C_{\mu_1^1\cdots \mu_{r_1}^1;\nu_1^1\cdots }
    =
    \sum_{\mathbb Y\in Y_{\rm DDI}} 
    d_{\mathbb Y}\,
     \eta_{\,\mathbb Y}
     +\sum_i\,d_{\bar a}\,\bar D^{\bar a}_{\mu_1^1\cdots \mu_{r_1}^1;\nu_1^1\cdots }\,.
     \label{n3 DDI}
\ee 
Compared to the $n\ge 4$ analog \eqref{DDI coupling}, 
we appended the terms involving $\bar D^{\bar a}_{\mu_1^1\cdots \mu_{r_1}^1;\nu_1^1\cdots }$ to incorporate the on-shell condition $w^{12}\approx 0$\,.
We examine whether
the term $\sum_{\mathbb Y\in Y_{\rm DDI}} 
d_{\mathbb Y}\,
\eta_{\,\mathbb Y}$ in the left hand side of the equation
can be removed by appropriately subtracting a 
coupling tensor $\Delta C_{\mu_1^1\cdots \mu_{r_1}^1;\nu_1^1\cdots }$ which is a DDI by itself:
 \be
      \Delta C_{\mu_1^1\cdots \mu_{r_1}^1;\nu_1^1\cdots}
      =\sum_{\mathbb Y\in Y_{\rm DDI}}
      \tilde d_{\mathbb Y}\,\eta_{\,\mathbb Y}\,,
      \qquad
      \mathsf S\,\Delta C_{\mu_1^1\cdots \mu_{r_1}^1;\nu_1^1\cdots} 
      = \sum_{\mathbb Y\in Y_{\rm DDI}} 
    d_{\mathbb Y}\,
     \eta_{\,\mathbb Y}\,.
\ee 
Differently from the $n\ge 4$ case, the property \eqref{S ddi} does not hold in this case, 
hence there is no universal way to get $\Delta C_{\mu_1^1\cdots \mu_{r_1}^1;\nu_1^1\cdots}$
for each $C_{\mu_1^1\cdots \mu_{r_1}^1;\nu_1^1\cdots}$
satisfying \eqref{n3 DDI},
and there are cases where $\Delta C_{\mu_1^1\cdots \mu_{r_1}^1;\nu_1^1\cdots}$
does not exist
and these cases correspond precisely
to the couplings whose gauge invariance rely on DDI.\footnote{One may wonder the property $\mathsf S\left(
 \sum d_{\mathbb Y}\,\eta_{\,\mathbb Y}
 +\sum d_{\bar a} \bar D^{\bar a}\right)
 =\sum d_{\mathbb Y}\,
 \eta_{\,\mathbb Y}
 +\sum d_{\bar a} \bar D^{\bar a}$ 
might be useful in this argument.
But, it just tells that $\sum d_{\bar a} \bar D^{\bar a}$
can be viewed as the coupling gauge invariant up to DDI.
Remember that $\bar D^{\bar a}$ is a linear 
combination of $B^m$
and $D^a$,
so $\sum d_{\bar a} \bar D^{\bar a}$ is a 
non-trivial on-shell coupling.}

To summarise, we have argued that for $n>3$ DDIs do not introduce new non-trivial solutions on top of the homogeneous equation solutions \eqref{n pt}. The effect of DDIs for $n>3$ is therefore relegated to produce linear relations between a priori different homogeneous solutions. For example in $d=3$ all Weyl tensor vanish for $l\geq2$ and the corresponding n-pt structure proportional to higher-spin fields therefore also vanishes. We conclude that in $d=3$ there cannot exist a non-vanishing parity preserving homogeneous solution with $n>3$ encoding a coupling of a massless higher-spin field to any massive/massless field of any spin, including scalars and spin-1. 

\subsection{Combining solutions: more than one massless field}

So far we have focused for simplicity on the solution space for couplings 
which is gauge invariant with respect to a single external leg. When more than one field is massless, one should then consider the intersection of the corresponding solution spaces associated with each legs (see e.g. \cite{Joung:2012hz} for a detailed discussion of this at the 3pt level). This is straightforward to analyse
 because we have shown that
 the gauge invariance of a coupling with respect to one field, say $\phi(x_i,u_i)$, allows the $\partial_{u_i}$ dependence only through
 the form of the curl $(\partial_{u_i})^{[\mu}\,(\partial_{x_i})^{\n]}$.
 For instance, when all the fields are massless, 
 all $\partial_{u_i}$ with $i=1,\ldots,n$ should enter
 through the curls. Then, the only allowed structures are
 the contractions between two curls:
 \be
    H_{ij}=2\,(\partial_{u_i})_{[\mu}\,(\partial_{x_i})_{\n]}
    \,(\partial_{u_j})^{[\mu}\,(\partial_{x_j})^{\n]}
    =z_{ij}\,w^{ij}-y^i_j\,y^j_i\,,
    \label{H ij}
 \ee
the contractions between one curl and $(\partial_{x_j})_\m\,(\partial_{x_k})_\n$\,:
\be
    H_i^{jk}=2\,(\partial_{u_i})_{[\mu}\,(\partial_{x_i})_{\n]}\,
    (x_j)^\m\,(x_k)^\n
    =y_i^j\,w^{ik}-y_i^k\,w^{ij}\,,
    \label{H ijk}
\ee
and the contractions of $\partial_{x_i}$'s, that is $w^{ij}$'s.
 In the end, the solution space 
 for the gauge invariant coupling of $n$ massless fields
 is
\be
	C^{\sst (n)}=K^{\sst (n)}\big(H_{ij}, H_i^{jk}, w^{ij}\big)\,.
	\label{n sol}
\ee
Again, $H_i^{jk}$ and $w^{ij}$ are not all independent but
satisfy
\be
    \sum_{j=1}^n H_i^{jk}\sim 0\,,\qquad \sum_{j=1}^n\,w^{ij}\sim 0\,.
\ee
An interesting observation that follows from the analysis of gauge invariant solutions we have done here is that all solutions found are in one-to-one correspondence with tensorial structures which can all be obtained from powers of spin-1 and scalars couplings.\footnote{This observation was made at cubic order in \cite{Sagnotti:2010at} and extended to higher point amplitudes in \cite{Taronna:2011kt} where it was also shown to be valid in the case of fermionic couplings. It can be considered as a generalisation of the double-copy structure of spinning homogeneous solutions.} This simply follows in this context from the fact that each of $H_{ij}$ and $H_i^{jk}$ are in one-to-one correspondence with gauge boson coupling to scalar field. 
Therefore this gives a different argument for the result obtained by a brute force analysis in \cite{Taronna:2011kt}.

In the following we will consider the generalisation of the flat-space analysis to (A)dS.

\section{Homogeneous Solution in AdS}\label{sec:AdS}

The analysis of the homogeneous couplings in flat space
can be straightforwardly generalised to the (A)dS case 
using the ambient space formalism:
$(A)dS_D$ space is the hypersurface $X^2=-L^2$ embedded in 
the ambient space $\mathbb{R}^{D+1}$,
and
the (A)dS fields are homogeneous,
$X\cdot\partial_X-U\cdot \partial_U+2-\mu)\Phi(X,U)=0$,
and tangent, $(X\cdot \partial_U\,\Phi(X,U)=0$. 
The starting point is again to express the homogeneous coupling $S^{\sst (n)}_{h}$ 
through a function $C_h$ as
\be
	S^{\sst (n)}_{h} \ett \int_{\rm\sst (A)dS}
	C_{h}^{\sst (n)}(W,Y,Z)\,\Phi_{1}(X_{1},U_{1})\,\cdots\,
	\Phi_{n}(X_{n},U_{n})\,\Big|_{{}^{X_{i}=X}_{U_{i}=0}}\,,
	\label{n homo}
\ee
where the variables $W^{ij}, Y^{j}_{i}$ and $Z_{ij}$ are 
the ambient space analogues of \eqref{vars}:
\be
	W^{ij}=\partial_{X_{i}}\cdot\partial_{X_{j}}\,,\qquad
	Y^{j}_{i}=\partial_{U_{i}}\cdot\partial_{X_{j}}\,,\qquad
	Z_{ij}=\partial_{U_{i}}\cdot\partial_{U_{j}}\,.
\ee
Similarly to the flat space cases, 
$Y^i_i$ and $Z_{ii}$ are excluded by the transversality and tracelessness
while $W^{ii}$ is removed by a field redefinition.
However, the flat-space relations \eqref{ToZeroRel},
given by the ambiguities of integration-by-parts,
are now modified into the following constraints imposed on $C_{h}^{\sst (n)}$\,:
\be \label{eqrelAdS}
    \cD^{Y}_i\,C^{\sst (n)}_{h}(W,Y,Z)=0\,,
    \qquad 
    \cD^{W}_i\,C^{\sst (n)}_{h}(W,Y,Z)=0\,,
\ee 
where $\cD^{Y}_i$ and $\cD^W_i$ are the differential operators,
\ba
	&&\cD^{Y}_i=\sum_{k=1}^{n}\left[ Y^{k}_{i}+\l
    \left(Z_{ik}\,\partial_{Y^{i}_{k}}+Y^{k}_{i}\,\partial_{W^{ik}}\right)\right],\\
	&& \cD^{W}_i=\sum_{k=1}^{n}\left[W^{ik}+\l
    \left(2-Z_{ik}\,\partial_{Z_{ik}}+Y^{k}_{i}\,\partial_{Y^{k}_{i}}
    -Y^i_k\,\partial_{Y^i_k}+
	W^{ik}\,\partial_{W^{ik}}\right)\right].
	\label{YW}
\ea
Here, $\lambda$ is defined as 
\be 
    \int_{\rm\sst (A)dS}\,\l^n=\int d^{D+1}X\,\delta^{[n]}(X^2-L^2)\,.
\ee
The operators $\cD^{Y}_i$ and $\cD^W_j$ commute among themselves and with each others.
This property is manifest if one considers their origin,
\be
    \cD^{Y}_i\sim \partial_{U_i}\cdot\left( \sum_{k=1}\partial_{X_k}\right),
    \qquad 
    \cD^{W}_i\sim \partial_{X_i}\cdot\left( \sum_{k=1}\partial_{X_k}\right).
    \label{DY DW origin}
\ee 
Therefore,
the numbers of \emph{independent} 
variables $W,Y,Z$\,, is obviously the same as in flat space:
\be
	\big(\,\#\ {\rm of}\ W\,\big)=\tfrac12\,n(n-3)\,,\quad
	\big(\,\#\ {\rm of}\ Y\,\big)=n(n-2)\,,\quad
	\big(\,\#\ {\rm of}\ Z\,\big)=\tfrac12\,n(n-1)\,.
	\label{num WYZ}
\ee
The solutions $C_{h}^{\sst (n)}$ of the above equations are in one-to-one correspondence to the TT part of all possible $n$-th order vertices.

The next step is to deal with the condition of gauge invariance \mt{\delta^{\sst (0)}S^{\sst (n)}_{h}\approx 0}\, in (A)dS.
The lowest order gauge symmetry of (A)dS fields
can be written in the ambient space formulation as $\delta\,\Phi(X,U)=U\cdot\partial_X\,E(X,U)$.
Therefore,
the gauge invariance of \eqref{n homo} 
up to an on-shell and TT conditions gives 
\be
	[\,C^{\sst (n)}(W,Y,Z)\,,\,U_{i}\cdot\partial_{X_{i}}\,]
	\approx 
	\cD_i^{\rm gauge}\, C^{\sst (n)}_{h}(W,Y,Z)=0\,.
	   	\label{n eq}
\ee	
where
\be
	\cD_i^{\rm gauge}=
	\sum_{k=1}^n\left(Y^{i}_{k}\,\partial_{Z_{ik}}+W^{ik}\,\partial_{Y^{k}_{i}}\right).
\ee
Notice here that we have treated the $Y$ and $W$ variables
as if they are all independent.
This is consistent because the differential operator $\cD_i^{\rm gauge}$ in eq.~\eqref{n eq} leaves the constrained surface given by eq.~\eqref{YW} invariant.\footnote{This is another way of saying that as in flat space case the characteristics of the above equations are at most linear regardless the choice of representative for the equivalence relation \eqref{eqrelAdS}.} 
This can be seen from the relations,
\be 
    [\,\cD_i^{\rm gauge}\,,\,\cD_j^{Y}\,]
    =\delta_{ij}\,\cD_i^W\,,
    \qquad
    [\,\cD_i^{\rm gauge}\,,\,\cD_j^{W}\,]=0\,,
\ee
which are again obvious consequences of \eqref{DY DW origin}.
Therefore, it is consistent to work with linearly dependent set of variables like we did in the flat-space analysis.
The simple structure of the PDE \eqref{n eq} which is basically the same as the flat space equation makes it possible
to identify its general solution precisely as in the flat space case:\footnote{Indeed the characteristics of the above equation are the same as in flat-space for $n>3$ since also in (A)dS it is always possible to fix a representative of the equivalence relation such that all characteristics are either linear or constant.}
\be
	C^{\sst (n)}=K^{\sst (n)}\big(H_{ij},H_i^{jk},W^{ij}\big)\,.
	\label{n sol AdS}
\ee
where $H_{ij}$ and $H^{jk}_{i}$ are defined by\footnote{We are focusing here for brevity directly on couplings of massless higher-spin field for which we must impose gauge invariance with respect to all external legs. In the case in which some of the field are massive one should also consider the AdS covariantisation of the additional $h$-structures we found in the flat space classification. This is straightforward in the ambient space.}
\be
	H_{ij}=Z_{ij}\,W^{ij}-Y^{i}_{j}\,Y^{j}_{i}\,,\qquad
	H^{jk}_{i}=Y^{j}_{i}\,W^{ik}-Y^{k}_{i}\,W^{ij}\,.
\ee
It is interesting to observe how one might have obtained the above solution by a straightforward covariantisation of the flat-space solutions \eqref{H ij} and \eqref{H ijk}. It is indeed very easy to argue in general (see e.g. \cite{Boulanger:2008tg,Joung:2011ww}) that for each (A)dS solution to the gauge invariance condition there exist a corresponding flat space solution in the limit $\Lambda\to0$. 
The number of flat-space solution is therefore an upper bound to the number of (A)dS solutions. Since all flat-space $h$-solutions \eqref{n pt}
can be trivially covariantised  to the (A)dS ones, we can conclude that the (A)dS $h$-solutions also form a complete set
for $n\ge 4$. 

Only the $n=3$ case is different.
Since some of $y_i^j$ variables  are gauge invariant on-shell,
there exist more solutions besides 
the $h$-structures.
To find out all solutions, 
we first solve the linear dependency
\eqref{YW} by discarding half of the $Y^{j}_{i}$'s and all the $W^{ij}$'s.
This will deform the equation \eqref{n eq} into
\be
	\cD_1^{\rm gauge}=
Y^{1}_{3}\,\partial_{Z_{31}}
-Y^3_2\,\partial_{Z_{12}}
+\lambda\left(Y_3^1\,\partial_{Y_3^1}
-Y_2^3\,\partial_{Y_2^3}\right)\,\partial_{Y_1^2}\,.
\ee
whose solutions were analyzed in \cite{Joung:2011ww,Joung:2012rv,Joung:2012fv,Taronna:2012gb}
and provide all gauge invariant $3$-pt couplings.\footnote{The radial reduction of these vertices has been also studied in \cite{Taronna:2012gb,Manvelyan:2012ww,Sleight:2016dba,Francia:2016weg,Karapetyan:2019psg}.}
Since the above solutions cover the entire solution space, 
the $h$-solutions should belong to it:
some solutions of the above equation
can be re-expressed as $h$-solutions (see e.g. \cite{Joung:2012hz} for additional details).

\section{Particular solutions, current exchanges \& locality}\label{sec:partSol}

Before concluding this note, we would like to spend a few lines by reviewing what is known about the particular solution to the Noether procedure \cite{Taronna:2011kt,Sleight:2017pcz}.

From the standard Noether procedure point of view, one is usually asked to directly find, under the locality assumption, a local particular solution. However this strategy proves to hide some conceptual simplicity. In the following, along the lines of \cite{Taronna:2011kt}, we shall not ask any further requirement on the particular solution with the aim of simplifying its structure relating it to the lower point couplings. The questions about locality will be raised at a different but equivalent level in terms of the full coupling defined above as\footnote{In \cite{Sleight:2017pcz} locality was replaced by the condition that the homogeneous solutions form representations of the HS algebra. This allows to uniquely fix the homogeneous solutions on-shell. Locality remains an important condition since the off-shell field frame cannot be fixed in this way and leads to divergences.}
\be
S^{\sst (n)}=S^{\sst (n)}_{h}+S^{\sst (n)}_{p}\,,
\ee
even though in the present letter we do not aim 
for this issue (we refer to \cite{Taronna:2011kt,Taronna:2012gb,Sleight:2017pcz}).
Since the particular solution is always defined up to a homogeneous one,
one can choose \emph{any} solution of
\be
\delta^{\sst (0)}S^{\sst (n)}_{p}
	\approx-\left(\delta^{\sst (1)}S^{\sst (n-1)}+\,\cdots\,
	+\delta^{\sst (n-3)} S^{\sst (3)}\right)\,.
\ee
Actually here exists a simple solution of the above equation --- the $n$-point amplitude composed by all the $m$-point vertices with $m\le n-1$\,.

To explain the point, let us use the diagramms
where
the quadratic and cubic actions are depicted as
\be 
    S^{\sst (2)}(\Phi_i)\,=\,
	\parbox{30mm}{\centering\begin{tikzpicture}
	\draw[>-<,thick] (-0.9,0) -- (0.9,0);
	\draw[thick, fill=black] (0,0) circle (0.1);
	\node [left]  at (-0.9,0) {$i$};
	\node [right] at (0.9,0) {$i$};
	\end{tikzpicture}},
	\qquad 
	 S^{\sst (3)}(\Phi_i,\Phi_j,\Phi_k)\,=\,
	\parbox{23mm}{\centering\begin{tikzpicture}
	\draw[>-,thick] (-0.5,-0.7) -- (0,0);
	\draw[-<,thick] (0,0) -- (0.9,0);
	\draw[-<,thick] (0,0) -- (-0.5,0.7);
	\node[above] at (-0.5,0.7)  {$i$};
	\node [below]  at (-0.5,-0.7) {$k$};
	\node [right] at (0.9,0) {$j$};
	\draw[thick, fill=black] (0,0) circle (0.1);
	\end{tikzpicture}},
\ee
and the lowest order gauge transformation  as
\be
	\delta^{\sst (0)}_i\Phi_i\,=\,
	\parbox{30mm}{\centering\begin{tikzpicture}
	\draw[>-,thick, dotted, blue] (-0.9,0) -- (0,0);
	\draw[->,thick] (0,0) -- (0.9,0);
	\draw[thick, fill=yellow] (0,0) circle (0.1);
	\node [left]  at (-0.9,0) {$i$};
	\node [right] at (0.9,0) {$i$};
	\end{tikzpicture}}.
\ee
The relation $\delta^{\sst (0)} S^{\sst (3)}+\delta^{\sst (1)}S^{\sst (2)}=0$, or more precisely,
\be
     \delta^{\sst (0)}_1\Phi_1\,\frac{\delta S^{\sst (3)}(\Phi_1,\Phi_2,\Phi_3)}{\delta \Phi_1}
     +\left[\delta^{\sst (1)}_1\Phi_2\right]_3\,
     \frac{\delta S^{\sst (2)}(\Phi_2)}{\delta \Phi_2}
     +\left[\delta^{\sst (1)}_1\Phi_3\right]_2\,
     \frac{\delta S^{\sst (2)}(\Phi_3)}{\delta \Phi_3}
     =0\,,
\ee
can be expressed diagrammatically as
\be
	\parbox{27mm}{\centering\begin{tikzpicture}
	\draw[>-,thick, dotted, blue] (-0.7,0.9) -- (-0.35,0.45);
	\draw[thick] (-0.35,0.45) -- (0,0);
	\draw[thick, fill=yellow] (-0.35,0.45) circle (0.1);
	\draw[-<,thick] (0,0) -- (1.1,0);
	\draw[-<,thick] (0,0) -- (-0.7,-0.9);
	\node[above] at (-0.7,0.9)  {$1$};
	\node [below]  at (-0.7,-0.9) {$2$};
	\node [right] at (1.1,0) {$3$};
	\draw[thick, fill=black] (0,0) circle (0.1);
	\end{tikzpicture}}
	+
	\parbox{27mm}{\centering\begin{tikzpicture}
	\draw[>-,thick, dotted, blue] (-0.7,0.9) -- (0,0);
	\draw[-<,thick] (0,0) -- (1.1,0);
	\draw[-<,thick] (0,0) -- (-0.7,-0.9);
	\node[above] at (-0.7,0.9)  {$1$};
	\node [below]  at (-0.7,-0.9) {$2$};
	\node [right] at (1.1,0) {$3$};
	\draw[thick, fill=yellow] (0,0) circle (0.1);
	\draw[thick, fill=black] (0.55,0) circle (0.1);
	\end{tikzpicture}}
	+
    \parbox{27mm}{\centering\begin{tikzpicture}
	\draw[>-,thick, dotted, blue] (-0.7,0.9) -- (0,0);
	\draw[-<,thick] (0,0) -- (1.1,0);
	\draw[-<,thick] (0,0) -- (-0.7,-0.9);
	\node[above] at (-0.7,0.9)  {$1$};
	\node [below]  at (-0.7,-0.9) {$2$};
	\node [right] at (1.1,0) {$3$};
	\draw[thick, fill=yellow] (0,0) circle (0.1);
	\draw[thick, fill=black] (-0.35,-.45) circle (0.1);
	\end{tikzpicture}}
	=0\,,
\ee
with $\left[\delta^{\sst (1)}_i\Phi_j\right]_k$
--- the part of the variation of the field $\Phi_j$ under the gauge
symmetry of $i$-th field 
which is linear in $\Phi_k$ --- depicted as
\be
    \left[\delta^{\sst (1)}_i\Phi_j\right]_k\,=\,
	\parbox{27mm}{\centering\begin{tikzpicture}
	\draw[>-,thick, dotted, blue] (-0.7,0.9) -- (0,0);
	\draw[->,thick] (0,0) -- (1.1,0);
	\draw[-<,thick] (0,0) -- (-0.7,-0.9);
	\node[above] at (-0.7,0.9)  {$i$};
	\node [below]  at (-0.7,-0.9) {$k$};
	\node [right] at (1.1,0) {$j$};
	\draw[thick, fill=yellow] (0,0) circle (0.1);
	\end{tikzpicture}}.
\ee
Here, we used the rule,
\be
    \parbox{27mm}{\centering\begin{tikzpicture}
	\draw[->,thick] (-1.2,0) -- (-0.2,0);
	\draw[>-,thick] (0.2,0) -- (1.2,0);
	\node at (0,0) {$i$};
	\end{tikzpicture}}
	=
	 \parbox{30mm}{\centering\begin{tikzpicture}
	\draw[-,thick] (-1,0) -- (1,0);
	\node[left] at (-1,0) {$i$};
	\node[right] at (1,0) {$i$};
	\end{tikzpicture}}.
\ee
Now we are ready to proceed to the 4-pt coupling
whose particular solution
can be chosen as the current-exchange part of the quartic amplitude:
\ba
	S^{\sst (4)}_{p}(\Phi_1,\Phi_2,\Phi_3,\Phi_4)
	\eq \parbox{30mm}{\centering\begin{tikzpicture}
	\draw[>-<, thick] (0,0.7) -- (2,-0.7);
	\draw[>-<,thick] (0,-0.7) -- (2,0.7);
	\draw[thick, fill=white] (1,0) circle (0.2);
	\node at (1,0) {$p$};
	\node[above] at (0,0.7) {$1$};
	\node[above] at (2,0.7) {$2$};
	\node[below] at (2,-0.7) {$3$};
	\node[below] at (0,-0.7) {$4$};
	\end{tikzpicture}}\nn
	\eq 
    \sum_{k}\ 
	\parbox{30mm}{\centering\begin{tikzpicture}
	\draw[>-<, thick] (0,.7) -- (.5,0) -- (0,-.7);
	\draw[>-<,thick] (2.3,.7) -- (1.8,0) -- (2.3,-.7);
	\draw[thick] (.5,0) -- (1.8,0);
	\draw[thick, fill=black] (0.5,0) circle (0.1);
	\draw[thick, fill=black] (1.8,0) circle (0.1);
	\draw[thick, fill=white] (1.15,0) circle (0.1);
	\node[below] at (1.15,0) {$k$};
	\node[above] at (0,0.7) {$1$};
	\node[above] at (2.3,0.7) {$2$};
	\node[below] at (2.3,-0.7) {$3$};
	\node[below] at (0,-0.7) {$4$};
	\end{tikzpicture}}
	+
	\parbox{25mm}{\centering\begin{tikzpicture}
	\draw[>-<,thick] (0.7,0) -- (0,0.5) -- (-0.7,0);
	\draw[>-<,thick] (0.7,2.3) -- (0,1.8) -- (-0.7,2.3);
	\draw[thick] (0,0.5) -- (0,1.8);
	\draw[thick, fill=black] (0,0.5) circle (0.1);
	\draw[thick, fill=black] (0,1.8) circle (0.1);
	\draw[thick, fill=white] (0,1.15) circle (0.1);
	\node[right] at (0,1.15) {$k$};
	\node[above] at (-0.7,2.3) {$1$};
	\node[above] at (0.7,2.3) {$2$};
	\node[below] at (0.7,0) {$3$};
	\node[below] at (-0.7,0) {$4$};
	\end{tikzpicture}}
	+
	\parbox{30mm}{\centering\begin{tikzpicture}
	\draw[>-<,thick] (1,0) -- (0.6,0.6) -- (-0.6,0.6) -- (-1,0);
	\draw[-<,thick] (0.6,0.6) -- (-1,1.6);
	\draw[-<,thick] (-0.6,0.6) -- (1,1.6);
	\draw[thick, fill=black] (0.6,0.6) circle (0.1);
	\draw[thick, fill=black] (-0.6,0.6) circle (0.1);
    \draw[thick, fill=white] (0,0.6) circle (0.1);	
	\node[below] at (0,0.6) {$k$};
	\node[above] at (-1,1.6) {$1$};
	\node[above] at (1,1.6) {$2$};
	\node[below] at (1,0) {$3$};
	\node[below] at (-1,0) {$4$};
	\end{tikzpicture}}.
	\label{part sol}
\ea
where the white circle in the internal line is the $1/\Box$ propagator for
the field $k$, hence satisfying
\be
    \parbox{30mm}{\centering\begin{tikzpicture}
	\draw[-,thick] (0,0) -- (2.1,0);
	\draw[thick, fill=black] (0.7,0) circle (0.1);
	\draw[thick, fill=white] (1.4,0) circle (0.1);
	\end{tikzpicture}}
	=
	 \parbox{30mm}{\centering\begin{tikzpicture}
	\draw[-,thick] (-1,0) -- (1,0);
	\end{tikzpicture}}
	+
	 \parbox{30mm}{\centering\begin{tikzpicture}
	\draw[-,thick] (-1,0) -- (-0.6,0);
	\draw[dotted,thick] (-0.6,0) -- (0.6,0);
	\draw[-,thick] (0.6,0) -- (1,0);
	\draw[thick, fill=white] (0,0) circle (0.1);
	\draw[thick, fill=yellow] (-0.6,0) circle (0.1);
	\draw[thick, fill=yellow] (0.6,0) circle (0.1);
	\end{tikzpicture}}.
	\label{prop}
\ee
The second term of the right hand side of the equality
indicates the divergence term.
The gauge variation of $S^{\sst (4)}_p$ \eqref{part sol} gives
\ba
 &&\parbox{30mm}{\centering\begin{tikzpicture}
	\draw[>-,thick, dotted, blue] (0,0.7) -- (0.5,0.35);
	\draw[-<, thick] (0.5,0.35) -- (2,-0.7);
	\draw[thick, fill=yellow] (0.5,0.35) circle (0.1);	
	\draw[>-<,thick] (0,-0.7) -- (2,0.7);
	\draw[thick, fill=white] (1,0) circle (0.2);
	\node at (1,0) {$p$};
	\node[above] at (0,0.7) {$1$};
	\node[above] at (2,0.7) {$2$};
	\node[below] at (2,-0.7) {$3$};
	\node[below] at (0,-0.7) {$4$};
	\end{tikzpicture}}
	=
    -\sum_{k}\ 
	\parbox{30mm}{\centering\begin{tikzpicture}
	\draw[>-,dotted, blue,  thick] (0,.7) -- (.5,0);
	\draw[-<, thick] (.5,0) -- (0,-.7);
	\draw[>-<,thick] (2.3,.7) -- (1.8,0) -- (2.3,-.7);
	\draw[thick] (.5,0) -- (1.8,0);
	\draw[thick, fill=yellow] (0.5,0) circle (0.1);
	\draw[thick, fill=black] (1.8,0) circle (0.1);
	\draw[thick, fill=white] (1.15,0) circle (0.1);
	\draw[thick, fill=black] (0.25,-0.35) circle (0.1);
	\node[below] at (1.15,0) {$k$};
	\node[above] at (0,0.7) {$1$};
	\node[above] at (2.3,0.7) {$2$};
	\node[below] at (2.3,-0.7) {$3$};
	\node[below] at (0,-0.7) {$4$};
	\end{tikzpicture}}
	+
	\parbox{30mm}{\centering\begin{tikzpicture}
	\draw[>-, thick, dotted, blue] (0,.7) -- (.5,0);
	\draw[-<, thick] (.5,0) -- (0,-.7);
	\draw[>-<,thick] (2.3,.7) -- (1.8,0) -- (2.3,-.7);
	\draw[thick] (.5,0) -- (1.8,0);
	\draw[thick, fill=yellow] (0.5,0) circle (0.1);
	\draw[thick, fill=black] (1.8,0) circle (0.1);
	\node[below] at (1.15,0) {$k$};
	\node[above] at (0,0.7) {$1$};
	\node[above] at (2.3,0.7) {$2$};
	\node[below] at (2.3,-0.7) {$3$};
	\node[below] at (0,-0.7) {$4$};
	\end{tikzpicture}}
	\nn
	&&\quad +
	\parbox{30mm}{\centering\begin{tikzpicture}
	\draw[>-,dotted, blue,  thick] (0,.7) -- (.5,0);
	\draw[-<, thick] (.5,0) -- (0,-.7);
	\draw[>-<,thick] (2.5,.7) -- (2,0) -- (2.5,-.7);
	\draw[dotted, thick] (.5,0) -- (2,0);
	\draw[thick, fill=yellow] (0.5,0) circle (0.1);
	\draw[thick, fill=yellow] (0.7,0) circle (0.1);
	\draw[thick, fill=yellow] (2,0) circle (0.1);
	\draw[thick, fill=white] (1.35,0) circle (0.1);
	\draw[thick, fill=black] (2.25,0.35) circle (0.1);
	\node[below] at (1.35,0) {$k$};
	\node[above] at (0,0.7) {$1$};
	\node[above] at (2.5,0.7) {$2$};
	\node[below] at (2.5,-0.7) {$3$};
	\node[below] at (0,-0.7) {$4$};
	\end{tikzpicture}}
	+
		\parbox{30mm}{\centering\begin{tikzpicture}
	\draw[>-,dotted, blue,  thick] (0,.7) -- (.5,0);
	\draw[-<, thick] (.5,0) -- (0,-.7);
	\draw[>-<,thick] (2.5,.7) -- (2,0) -- (2.5,-.7);
	\draw[dotted, thick] (.5,0) -- (2,0);
	\draw[thick, fill=yellow] (0.5,0) circle (0.1);
	\draw[thick, fill=yellow] (0.7,0) circle (0.1);
	\draw[thick, fill=yellow] (2,0) circle (0.1);
	\draw[thick, fill=white] (1.35,0) circle (0.1);
	\draw[thick, fill=black] (2.25,-0.35) circle (0.1);
	\node[below] at (1.35,0) {$k$};
	\node[above] at (0,0.7) {$1$};
	\node[above] at (2.5,0.7) {$2$};
	\node[below] at (2.5,-0.7) {$3$};
	\node[below] at (0,-0.7) {$4$};
	\end{tikzpicture}}
	+ (2\leftrightarrow 4)
	+(3\leftrightarrow 4).
\ea
where we used \eqref{prop} to get the diagrammes
with two bullets.
These diagrammes correspond to 
$\delta^{\sst (1)}S^{\sst (3)}$
while the diagrammes with more bullets
to $\delta^{\sst (2)}S^{\sst (2)}$ with
\be
	\left[\delta^{\sst (2)}_i\Phi_j\right]_{pq}\,=\,
	\sum_k
	\parbox{30mm}{\centering\begin{tikzpicture}
	\draw[>-,dotted, blue,  thick] (0,.7) -- (.5,0);
	\draw[->, thick] (.5,0) -- (0,-.7);
	\draw[>-<,thick] (2.3,.7) -- (1.8,0) -- (2.3,-.7);
	\draw[thick] (.5,0) -- (1.8,0);
	\draw[thick, fill=yellow] (0.5,0) circle (0.1);
	\draw[thick, fill=black] (1.8,0) circle (0.1);
	\draw[thick, fill=white] (1.15,0) circle (0.1);
	\node[below] at (1.15,0) {$k$};
	\node[above] at (0,0.7) {$i$};
	\node[above] at (2.3,0.7) {$p$};
	\node[below] at (2.3,-0.7) {$q$};
	\node[below] at (0,-0.7) {$j$};
	\end{tikzpicture}}
		+
	\parbox{25mm}{\centering\begin{tikzpicture}
	\draw[>->,thick] (0.7,0) -- (0,0.5) -- (-0.7,0);
	\draw[dotted,blue,thick,-<] (0,1.8) -- (-0.7,2.3);
	\draw[>-,thick] (0.7,2.3) -- (0,1.8) ;
	\draw[thick, dotted] (0,0.5) -- (0,1.8);
	\draw[thick, fill=yellow] (0,0.5) circle (0.1);
	\draw[thick, fill=yellow] (0,1.8) circle (0.1);
	\draw[thick, fill=yellow] (0,1.6) circle (0.1);
	\draw[thick, fill=white] (0,1.05) circle (0.1);
	\node[right] at (0,1.05) {$k$};
	\node[above] at (-0.7,2.3) {$i$};
	\node[above] at (0.7,2.3) {$p$};
	\node[below] at (0.7,0) {$q$};
	\node[below] at (-0.7,0) {$j$};
	\end{tikzpicture}}
	+
	\parbox{30mm}{\centering\begin{tikzpicture}
	\draw[>-,thick] (1,0) -- (0.6,0.6);
	\draw[dotted,thick] (0.6,0.6) -- (-0.6,0.6);
	\draw [->,thick] (-0.6,0.6)-- (-1,0);
	\draw[-<,thick,blue, dotted] (0.6,0.6) -- (-1,1.6);
	\draw[-<,thick] (-0.6,0.6) -- (1,1.6);
	\draw[thick, fill=yellow] (0.6,0.6) circle (0.1);
	\draw[thick, fill=yellow] (0.4,0.6) circle (0.1);
	\draw[thick, fill=yellow] (-0.6,0.6) circle (0.1);
    \draw[thick, fill=white] (-0.1,0.6) circle (0.1);	
	\node[below] at (-0.1,0.6) {$k$};
	\node[above] at (-1,1.6) {$i$};
	\node[above] at (1,1.6) {$p$};
	\node[below] at (1,0) {$q$};
	\node[below] at (-1,0) {$j$};
	\end{tikzpicture}}.
\ee
Notice that the above $\delta^{\sst (2)}$ is generically non-local.
Since we have shown that the homogeneous solution $S^{\sst (4)}_h$ does not induce any
$\delta^{\sst (2)}$, this seems in contradiction with existence of any local theory.
The point is that the latter statement holds only up to a field redefinition,
which may be non-local as well but well-defined on shell.
To summarise, a homogeneous solution $S_h^{\sst (4)}$ resulting 
from a non-local field redefinition,
\be
    \Phi=\parbox{20mm}{\centering\begin{tikzpicture}
	\draw[<-<, thick] (-.6,-.6) -- (0.6,.6);
	\end{tikzpicture}}
	 \longrightarrow   
	\parbox{20mm}{\centering\begin{tikzpicture}
	\draw[<-<, thick] (-.6,-.6) -- (0.6,.6);
	\end{tikzpicture}}
	+
     \parbox{25mm}{\centering\begin{tikzpicture}
	\draw[>-<,thick] (-0.8,0.8) -- (0.8,-0.8);
	\draw[<-<, thick] (-.8,-.8) -- (0.8,.8);
	\draw[thick, fill=green] (0,0) circle (0.2);
	\end{tikzpicture}},
\ee 
may compensate the non-local deformation of gauge transformation:
\be
    \parbox{25mm}{\centering\begin{tikzpicture}
	\draw[>-,dotted, blue,  thick] (-0.8,.8) -- (-0.45,0.45);
	\draw[-<,thick] (-0.45,0.45) -- (0.8,-0.8);
	\draw[<-<, thick] (-.8,-.8) -- (0.8,.8);
	\draw[thick, fill=yellow] (-0.45,0.45) circle (0.1);
	\draw[thick, fill=green] (0,0) circle (0.2);
	\end{tikzpicture}}=
	\parbox{30mm}{\centering\begin{tikzpicture}
	\draw[>-,dotted, blue,  thick] (0,.7) -- (.5,0);
	\draw[->, thick] (.5,0) -- (0,-.7);
	\draw[>-<,thick] (2.3,.7) -- (1.8,0) -- (2.3,-.7);
	\draw[thick] (.5,0) -- (1.8,0);
	\draw[thick, fill=yellow] (0.5,0) circle (0.1);
	\draw[thick, fill=black] (1.8,0) circle (0.1);
	\draw[thick, fill=white] (1.15,0) circle (0.1);
	\end{tikzpicture}}
		+
	\parbox{25mm}{\centering\begin{tikzpicture}
	\draw[>->,thick] (0.7,0) -- (0,0.5) -- (-0.7,0);
	\draw[dotted,blue,thick,-<] (0,1.8) -- (-0.7,2.3);
	\draw[>-,thick] (0.7,2.3) -- (0,1.8) ;
	\draw[thick, dotted] (0,0.5) -- (0,1.8);
	\draw[thick, fill=yellow] (0,0.5) circle (0.1);
	\draw[thick, fill=yellow] (0,1.8) circle (0.1);
	\draw[thick, fill=yellow] (0,1.6) circle (0.1);
	\draw[thick, fill=white] (0,1.05) circle (0.1);
	\end{tikzpicture}}
	+
	\parbox{30mm}{\centering\begin{tikzpicture}
	\draw[>-,thick] (1,0) -- (0.6,0.6);
	\draw[dotted,thick] (0.6,0.6) -- (-0.6,0.6);
	\draw [->,thick] (-0.6,0.6)-- (-1,0);
	\draw[-<,thick,blue, dotted] (0.6,0.6) -- (-1,1.6);
	\draw[-<,thick] (-0.6,0.6) -- (1,1.6);
	\draw[thick, fill=yellow] (0.6,0.6) circle (0.1);
	\draw[thick, fill=yellow] (0.4,0.6) circle (0.1);
	\draw[thick, fill=yellow] (-0.6,0.6) circle (0.1);
    \draw[thick, fill=white] (-0.1,0.6) circle (0.1);	
	\end{tikzpicture}}.
\ee
This is indeed what happens in all local theories.
Appendix \ref{app: YM} contains an explicit treatment of this point through 
the example of Yang-Mills theory and Einstein Gravity.
In the end, combining the particular solution $S^{\sst (4)}_p$ \eqref{part sol} 
with a homogeneous solution $S^{\sst (4)}_h$
--- consisting of a $h$-coupling \eqref{n sol}
and a non-local field redefinition ---
we obtain the most general form of consistent quartic couplings.
In local theories, the non-localities in $S_h^{\sst (4)}$
should cancel on-shell those in $S_p^{\sst (4)}$,
and this can happen only up to a suitable non-local but non-singular field redefinition.\footnote{We stress here that such redefinitions are in one to one correspondence with non-local functionals which vanish on-shell. This is different from non-local redefinitions which are expressed in terms of singular functionals and can remove non-trivial couplings. Most importantly such singular redefinition should not be allowed and pose a question for the off-shell definition of any $n$-point coupling in a theory with infinitely many fields (see e.g. the discussion in \cite{Taronna:2017wbx,Sleight:2017pcz}).}

For higher-order interactions, the particular solutions $S^{\sst (n)}_{p}$ can be found in the same manner as the quartic one, in terms of the homogeneous solutions $S^{\sst (m)}_{h}$ with $m\le n-1$\,. For instance, the quintic and the sextic couplings are given schematically by
\ba
	&&S^{\sst (5)}_{p}=\quad 
	\parbox{35mm}{\centering\begin{tikzpicture}
	\draw[>-<,thick] (0,.7) -- (.5,0) -- (0,-.7);
	\draw[>-<,thick] (3,.7) -- (2.5,0) -- (3,-.7);
	\draw[thick] (.5,0) -- (2.5,0);
	\draw[-<,thick] (1.5,0) -- (1.5,0.7);
	\draw[thick, fill=black] (0.5,0) circle (0.1);
	\draw[thick, fill=white] (1,0) circle (0.1);
	\draw[thick, fill=white] (2,0) circle (0.1);
	\draw[thick, fill=black] (2.5,0) circle (0.1);
	\draw[thick, fill=black] (1.5,0) circle (0.1);
	\end{tikzpicture}}
	+
	\parbox{30mm}{\centering\begin{tikzpicture}
	\draw[>-<,thick] (0,.7) -- (0.5,0) -- (0,-.7);
	\draw[-<,thick] (0.5,0) -- (2.2,0);
	\draw[>-<,thick] (1.5,-0.7) -- (1.5,0.7);
	\draw[thick, fill=white] (1.5,0) circle (0.2);
	\node at (1.5,0) {$h$};
	\draw[thick, fill=white] (1,0) circle (0.1);
	\draw[thick, fill=black] (.5,0) circle (0.1);
	\end{tikzpicture}},
	\\[10pt]
	&&S^{\sst (6)}_{p}=\quad
	\parbox{45mm}{\centering\begin{tikzpicture}
	\draw[>-<,thick] (0,.7) -- (.5,0) -- (0,-.7);
	\draw[>-<,thick] (4,.7) -- (3.5,0) -- (4,-.7);
	\draw[thick] (.5,0) -- (3.5,0);
	\draw[-<,thick] (1.5,0) -- (1.5,0.7);
	\draw[-<,thick] (2.5,0) -- (2.5,0.7);
	\draw[thick, fill=black] (0.5,0) circle (0.1);
	\draw[thick, fill=white] (1,0) circle (0.1);
	\draw[thick, fill=black] (1.5,0) circle (0.1);
	\draw[thick, fill=white] (2,0) circle (0.1);
	\draw[thick, fill=black] (2.5,0) circle (0.1);
	\draw[thick, fill=white] (3,0) circle (0.1);
	\draw[thick, fill=black] (3.5,0) circle (0.1);
	\end{tikzpicture}}
	+
	\parbox{35mm}{\centering\begin{tikzpicture}
	\draw[>-<,thick] (0,.7) -- (.5,0) -- (0,-.7);
	\draw[-<,thick] (0.5,0) -- (3.2,0);
	\draw[>-<,thick] (2.5,-0.7) -- (2.5,0.7);
	\draw[-<,thick] (1.5,0) -- (1.5,0.7);
	\draw[thick, fill=black] (0.5,0) circle (0.1);
	\draw[thick, fill=white] (1,0) circle (0.1);
	\draw[thick, fill=black] (1.5,0) circle (0.1);
	\draw[thick, fill=white] (2,0) circle (0.1);
	\draw[thick, fill=white] (2.5,0) circle (0.2);
    \node at (2.5,0) {$h$};
	\end{tikzpicture}}\nn
	&&
	\quad+ 
	\parbox{35mm}{\centering\begin{tikzpicture}
	\draw[>-<,thick] (0,.7) -- (.5,0) -- (0,-.7);
	\draw[>-<,thick] (3,.7) -- (2.5,0) -- (3,-.7);
	\draw[thick] (.5,0) -- (2.5,0);
	\draw[>-<,thick] (1.5,-0.7) -- (1.5,0.7);
	\draw[thick, fill=black] (.5,0) circle (0.1);	
	\draw[thick, fill=white] (1,0) circle (0.1);
	\draw[thick, fill=white] (1.5,0) circle (0.2);
	\node at (1.5,0) {$h$};
	\draw[thick, fill=white] (2,0) circle (0.1);	
	\draw[thick, fill=black] (2.5,0) circle (0.1);	
	\end{tikzpicture}}
    +
	\parbox{30mm}{\centering\begin{tikzpicture}
	\draw[>-<,thick] (.3,0) -- (2.7,0);
	\draw[>-<,thick] (1,-0.7) -- (1,0.7);
	\draw[>-<,thick] (2,-0.7) -- (2,0.7);
	\draw[thick, fill=white] (1,0) circle (0.2);
	\node at (1,0) {$h$};
	\draw[thick, fill=white] (1.5,0) circle (0.1);			\draw[thick, fill=white] (2,0) circle (0.2);
	\node at (2,0) {$h$};
	\end{tikzpicture}}
	+
	\parbox{30mm}{\centering\begin{tikzpicture}
	\draw[>-<,thick] (0,.7) -- (0.5,0) -- (0,-.7);
	\draw[>-<,thick] (2,.7) -- (1.5,0) -- (1,-.7);
	\draw[>-<,thick] (2,.-.7) -- (1.5,0) -- (1,.7);
	\draw[thick] (0.5,0) -- (1.5,0);
	\draw[thick, fill=black] (0.5,0) circle (0.1);
	\draw[thick, fill=white] (1,0) circle (0.1);
	\draw[thick, fill=white] (1.5,0) circle (0.2);
	\node at (1.5,0) {$h$};
	\end{tikzpicture}},
\ea
where the $h$-bullet vertices label the corresponding homogeneous solutions:
\be
	S^{\sst (4)}_{h}=\ 
	 \parbox{30mm}{\begin{tikzpicture}
	\draw[>-<,thick] (-0.7,-0.7) -- (0.7,0.7);
	\draw[>-<,thick] (-0.7,0.7) -- (0.7,-0.7);
	\draw[-<,thick, fill=white] (0,0) circle (0.2);
	\node at (0,0) {$h$};
	\end{tikzpicture}}\hspace{-35pt}\,,
	\qquad 
	S^{\sst (5)}_{h}=\
	 \parbox{30mm}{\begin{tikzpicture}
	 \draw[>-,thick] (0,0.9) -- (0,0);
	\draw[>-<,thick] (-0.6,-0.7) -- (0,0) -- (0.8,0.4);
	\draw[>-<,thick] (-0.8,0.4) -- (0,0) -- (0.6,-0.7);
	\draw[>-<,thick, fill=white] (0,0) circle (0.2);
	\node at (0,0) {$h$};
	\end{tikzpicture}}\hspace{-35pt}\,,
\ee
and the permutations of the external legs are suppressed for simplicity.
From these particular solutions, one can extract the corresponding deformation of gauge transformations, $\delta^{\sst (n-2)}$, which are generically non-local.
If the theory under consideration is local, then this is equivalent to the existence of a non-singular, non-local redefinition
which would remove all the non-local terms in $\delta^{\sst (n-2)}$.
If this field redefinition removes all terms 
without leaving any local remainders, 
the gauge algebra is at most linear in the fields,
which is the case in YM theory and Gravity.\footnote{Note that the deformations of the gauge transformations are linear only in the YM case while in the Gravity case they receive all orders deformations which resum to give the inverse metric. The gauge algebra is however fixed at the cubic level and does not receive further non-trivial corrections.}

Before adding more comments on the locality issue, let
us briefly remark that
the approach taken in this section did not assume the TT condition:
the cubic coupling $S^{\sst (3)}$ and the homogeneous 
solution $S^{\sst (n)}$ may
contain the divergence and trace terms. 
Even when we impose the on-shell conditions 
on the external legs, some of such non-TT terms survive
and contribute to the construction of $S^{\sst (m)}_p$. Such contribution is generically a contact contribution.

Let us make a few comments on the non-locality issues in
higher-spin(-like) theories.
As we explained above, if the sought theory is local, the 
homogeneous solution $S_h^{\sst (n)}$ --- consisting of 
a $h$-coupling and a non-local field redefinition  ---
can be uniquely determined
because we can use all our freedom of non-local field redefinitions
in removing the non-localities 
and this is possible only upon a unique choice of $h$-coupling.
On the contrary, in non-local theories,
it is, by definition, impossible to remove all the non-localities.
Therefore, we can neither fix the $h$-coupling nor the non-local field-redefinition:
we need a new criterion in order to avoid this ambiguity.
For instance, one can forbid the coupling to involve 
any singularities of physical exchange.
This condition allows for, besides local interactions, a vast range of non-local interactions. When there is a solution to this condition, 
it would fix, for instance, the $h$-coupling part (or the on-shell non-trivial part) of $S^{\sst (4)}_h$ in terms of the cubic coupling $S^{\sst (3)}$
(as shown in \cite{Taronna:2011kt,Taronna:2017wbx}).\footnote{It was also argued in \cite{Sleight:2017pcz} that 
massless higher spins in AdS would not admit such a solution, although it is possible to fix the on-shell part of the homogeneous solution via holography \cite{Bekaert:2014cea,Bekaert:2015tva,Sleight:2016dba,Sleight:2016hyl} or global higher-spin symmetry \cite{Sleight:2016xqq,Sleight:2017pcz}. The off-shell contribution would lead to divergences of the $1/\Box$ type associated to physical exchange contribution within the contact term, putting under question a purely field theoretic formulation of the theory.
}
The last but not the least important point 
relevant to the discussions of this paragraph is
the convergence issue in the sum $\sum_k$ in $S^{\sst (4)}_p$,
which may obscure even further the nature of the non-locality 
is $S^{\sst (4)}_p$ (see e.g. \S(4.3) of \cite{Taronna:2017wbx}).



We conclude this discussion stressing that a possibly interesting application of the results presented in this note is in the connection between the homogeneous solution to the Noether procedure in (A)dS and the $n$-point conformal structures of conserved currents. In this context gauge invariance in the bulk translates into current conservation on the boundary and this ambient space formulation could potentially be used in applications to CFT correlators with conserved spinning currents. We leave to the future a more detailed analysis of this connection and the possible applications to spinning correlators.

\acknowledgments

M.T. is grateful to Carlo Iazeolla and Per Sundell for useful discussions and especially to Charlotte Sleight for useful discussions and comments on the draft. This work was initiated at Scuola Normale Superiore in 2012, which we thank for support and hospitality.
The research of M.T. was partially supported by the European Union's Horizon 2020 research and innovation programme under the Marie Sklodowska-Curie grant agreement No 747228, by the program  “Rita  Levi  Montalcini”  of the MIUR (Minister for Instruction, University and Research) and by the INFN initiative STEFI.
The work of E.J. was supported by
National Research Foundation
(Korea)
through the grant
NRF-2019R1F1A1044065.

\appendix

\section{YM \& Gravity examples}\label{app: YM}

The aim of this Appendix is to give some examples of the general formalism and results which we have discussed in this note.

Focusing first on the YM case we shall start fom the cubic vertex
\begin{equation}
    \mathcal{C}^{\sst (3)}_{123}=g\,\left(\pl_{u_1}\cdot \pl_{x_{23}}\pl_{u_2}\cdot\pl_{u_3}+\pl_{u_2}\cdot \pl_{x_{31}}\pl_{u_3}\cdot\pl_{u_1}+\pl_{u_3}\cdot \pl_{x_{12}}\pl_{u_1}\cdot\pl_{u_2}\right)\,.
\end{equation}
For simplicity we already fixed the above color ordered form assuming that all coupling constants for different gauge fields are equal to $g$. It is also possible to consider different coupling constants to start with and use the procedure outlined below to fix them from the requirement of locality (see e.g. \cite{Taronna:2011kt,Taronna:2017wbx})! In the following for convenience we shall fix $g=1$ without loss of generality.
We can then easily evaluate the color ordered part of the exchange:
\begin{align}
    \mathcal{E}_{1234}=\mathcal{C}^{\sst (3)}_{12u}\,\mathcal{C}^{\sst (3)}_{34v}\,\frac{u\cdot v}{\mathsf{s}}+\mathcal{C}^{\sst(3)}_{41u}\,\mathcal{C}^{\sst(3)}_{23v}\,\frac{u\cdot v}{\mathsf{u}}\,.
\end{align}
Considering the gauge variation with respect to leg $1$ of the cubic structures one then gets:
\begin{equation}
    \delta^{\sst (0)}_{\epsilon_1}\mathcal{C}^{\sst (3)}_{12u}=(\pl_{x_2}^2-\pl_{x_u}^2)\,\pl_{u_2}\cdot\pl_u-\pl_{u}\cdot(\pl_{x_3}+\pl_{x_4})\,\pl_{u_2}\cdot(\pl_{x_3}+\pl_{x_4})+\pl_{u}\cdot\pl_{x_2}\,\pl_{u_2}\cdot\pl_{x_2}
\end{equation}
Working off-shell with respect to the external leg but factoring out terms proportional to divergences and traces we can then write
\begin{equation}
    \delta^{\sst (0)}_{\epsilon_1}\mathcal{C}^{\sst (3)}_{12u}=(\pl_{x_2}^2-\pl_{x_u}^2)\,\pl_{u_2}\cdot\pl_u-\pl_{u}\cdot(\pl_{x_3}+\pl_{x_4})\,\pl_{u_2}\cdot(\pl_{x_3}+\pl_{x_4})\,.
\end{equation}
The term proportional to $\pl_{x_u}^2$ compensates the propagator pole in the exchange, generating a local gauge variation. On the other hand the term proportional to $\pl_u\cdot(\pl_{x_3}+\pl_{x_4})$ computes the divergence of $\mathcal{V}_{34v}$ which is a conserved current on-shell, therefore generating further terms proportional to $\pl_{x_3}^2$ and $\pl_{x_4}^2$:
\begin{align}
    \pl_v\cdot(\pl_{x_3}+\pl_{x_4})\,\mathcal{C}^{\sst (3)}_{34v}
    =(\pl_{x_3}^2-\pl_{x_4}^2)\,\pl_{u_3}\cdot\pl_{u_4}
    +\pl_{u_3}\cdot\pl_{x_4}\,\pl_{u_4}\cdot\pl_{x_4}
    -\pl_{u_4}\cdot\pl_{x_3}\,\pl_{u_3}\cdot\pl_{x_3}\,.
\end{align}
It is interesting to notice that choosing a different representative for $\mathcal{V}_{12u}$ like the cyclic ansatz:
\begin{equation}
    \mathcal{C}^{\sst (3)}_{123}=g\,\left(\pl_{u_1}\cdot \pl_{x_{2}}\pl_{u_2}\cdot\pl_{u_3}+\pl_{u_2}\cdot \pl_{x_{3}}\pl_{u_3}\cdot\pl_{u_1}+\pl_{u_3}\cdot \pl_{x_{1}}\pl_{u_1}\cdot\pl_{u_2}\right)
\end{equation}
gives a gauge variation which precisely differ by such divergence terms. Consistency of the massless interaction ensures that such terms are proportional to the EoMs of the external legs. Gathering all such terms one obtains a rather cumbersome non-local deformation of the gauge transformations!

In the following we want to show that there exist a field and gauge parameter redefinition which removes the above non-local contributions to the gauge transformations. To this end, we proceed by solving for the homogeneous solution imposing locality of the quartic vertex on-shell. The most general ansatz for the homogeneous solution $\mathcal{C}^ {(4)}_{1234}=f(h_{ij},h_{i}^{jk},w^{ij})$ reads in this case:
\begin{align}\label{Eansatz}
    {\mathcal{C}}^{\sst (4)}_{1234}&=c_1\, h_{14}\, h_{23} + c_2\, h_{13}\, h_{24} 
    + c_3\, h_{12} \,h_{34} \nonumber\\&
    + c_4\, h_1^{34}\,h_2^{34}\, h_{34} + c_5\,h_1^{34}\, h_{24}\,h_3^{12} 
    + c_6 \,h_{14}\,h_2^{34}\,h_3^{12}\nonumber\\& + 
    c_7\,h_1^{34}\, h_{23}\, h_4^{12} 
    + c_8 \,h_{13} \,h_2^{34}\,h_4^{12} + c_9\, h_{12}\, h_3^{12}\,h_4^{12} \nonumber\\
    &+ c_{10}\,h_1^{34}\,h_2^{34}\,h_3^{12}\,h_4^{12}\,,
\end{align}
where we have fixed the on-shell redundancies in the $h$ monomials by picking the following ten independent structures:
\begin{align}
    &h_{12}\,,& &h_{34}\,,& &h_{13}\,,& &h_{24}\,,& &h_{14}\,,& &h_{23}\,,\nonumber\\
    && &h_1^{34}\,,& &h_2^{34}\,,& &h_3^{12}\,,& &h_4^{12}\,.
\end{align}
For instance only one among $h_1^{23}$, $h_1^{24}$ and $h_1^{34}$ is independent on-shell:
\begin{align}
    h_1^{23}-h_1^{34}&\approx0,& h_1^{24}+h_1^{34}&\approx0\,.
\end{align}
In the ansatz \eqref{Eansatz} we have also assumed that the coefficients of each structure are (possibly non-local) functions of $w^{ij}$.
The Noether procedure equation solves the quartic couplings as the following (generically non-local) functional:
\begin{equation}\label{Noether}
    \mathcal{V}_{1234}=\mathcal{C}^{\sst (4)}_{h\,1234}-{\mathcal{E}}_{1234}\,,
\end{equation}
Obviously, the Noether procedure equation above is empty unless further conditions on $\mathcal{V}_{1234}$ are enforced. Imposing locality of $\mathcal{V}_{1234}$ on-shell implies by dimensional analysis the following parameterisation for the quartic coupling:
\begin{equation}\label{localV}
    \mathcal{V}_{1234}\approx d_1\, \pl_{u_1}\cdot \pl_{u_2} \pl_{u_3}\cdot\pl_{u_4}+d_2\, \pl_{u_1}\cdot \pl_{u_3} \pl_{u_2}\cdot\pl_{u_4}+d_3\, \pl_{u_1}\cdot \pl_{u_4} \pl_{u_2}\cdot\pl_{u_3}\,,
\end{equation}
where the equality $\approx$ holds on-shell with respect to the external legs.
Note that by dimensional analysis a local vertex cannot be proportional in this case to $u_i\cdot p_j$.
We can now solve \eqref{localV}. We do this by first setting $\pl_{x_i}^2\approx 0$, $\pl_{x_i}\cdot \pl_{u_i}\approx 0$ and $\mathsf{t}\approx-\mathsf{s}-\mathsf{u}$ and integrating by parts all $\pl_{x_4}$ derivatives. After eliminating all on-shell redundancies eq.~\eqref{localV} becomes a simple linear equation for the coefficients $c_i$ and $d_i$. Its solution can be shown to be unique and reads for the $d_i$:
\begin{align}
    d_1&=-2\,,& d_2&=4\,,& d_3&=-2\,,
\end{align}
reproducing the YM quartic vertex in its color ordered form. The coefficients $c_i$ then read:
\begin{equation}
\begin{split}
c_1=\frac{4\,\mathsf{t}}{\mathsf{s}^3}\,,\qquad
c_2=\frac{4}{\mathsf{t}^2}\,,\qquad 
c_{3}=\frac{4\,\mathsf{t}}{\mathsf{u}^3}\,\,\qquad
c_4=c_{9}=\frac{4}{\mathsf{s}\,\mathsf{u}^3}\,,\\
c_6=c_7=\frac{4}{\mathsf{u}\,\mathsf{s}^3}\,,\qquad 
c_5=c_8=\frac{4}{\mathsf{s}\,\mathsf{u}\,\mathsf{t}^2}\,,\qquad 
c_{10}=-\frac{4(\mathsf{s}^2+\mathsf{s}\,\mathsf{u}+\mathsf{u}^2)}{\mathsf{s}^3\,\mathsf{u}^3\,\mathsf{t}^2}\,.
\end{split}
\end{equation}
The above choice of coefficients ensures that the contact term $\mathcal{V}_{1234}$ is local on-shell! However off-shell the contact term $\mathcal{V}_{1234}=\mathcal{C}^ {(4)}-{\mathcal{E}}_{1234}$ is actually non-local, which explains the non-local form of the deformations of the gauge transformations which can be obtained from the current exchange amplitude $\mathcal{E}_{1234}$, as discussed at the beginning of this section. However the fact that $\mathcal{V}_{1234}$ is local on-shell also implies that there exist a non-singular field redefinition which makes $\mathcal{V}_{1234}$ manifestly local. This implies that the non-local gauge deformations are in fact removable by a non-local non-singular field redefinition which maps off-shell $\mathcal{V}_{1234}$ to its manifestly local form \eqref{localV}. It is important to stress that the field redefinition that maps the two forms of the contact term is also non-local involving inverse of the Mandelstam invariants. This precisely accounts for the non-local deformation of the gauge transformation obtained from the exchange amplitude at the beginning of this section. The explicit form of the field redefinition is quite cumbersome and is not needed for our argument.

Similar but more cumbersome results can also be obtained in the case of massless spin-2 self-interactions. Skipping the details, after imposing locality on-shell of the solution $\mathcal{V}=\mathcal{C}^{(4)}-\mathcal{E}$ we find the following unique two-derivative quartic vertex:

{\allowdisplaybreaks\small
\begin{align}
    \mathcal{V}&=16 \Bigg[
    -2 (y_3^1) (y_4^2) (z_{12}) (z_{14}) (z_{23}) 
    -  (y_3^2) (y_4^2) (z_{12}) (z_{14}) (z_{23}) 
    -  (y_3^2) (y_4^3) (z_{12}) (z_{14}) (z_{23}) \nonumber\\&
    -  (y_2^3) (y_4^2) (z_{13}) (z_{14}) (z_{23}) 
    - 2 (y_2^1) (y_4^3) (z_{13}) (z_{14}) (z_{23})
    -  (y_2^3) (y_4^3) (z_{13}) (z_{14}) (z_{23}) \nonumber\\&
    - 2 (y_2^1) (y_3^1) (z_{14})^2 (z_{23}) 
    -  (y_2^3) (y_3^1) (z_{14})^2 (z_{23}) 
    -  (y_2^1) (y_3^2) (z_{14})^2 (z_{23})\nonumber\\&
    + (y_1^3) (y_4^2) (z_{14}) (z_{23})^2 
    + (y_1^2) (y_4^3) (z_{14}) (z_{23})^2 
    -  \tfrac{\mathsf{u}}{4}  (z_{14})^2 (z_{23})^2 \nonumber\\&
    + (y_3^1) (y_4^2) (z_{12}) (z_{13}) (z_{24}) 
    + 2 (y_3^2) (y_4^2) (z_{12}) (z_{13}) (z_{24}) 
    + 2 (y_3^2) (y_4^3) (z_{12}) (z_{13}) (z_{24}) \nonumber\\&
    -  (y_2^3) (y_4^2) (z_{13})^2 (z_{24})
    + (y_2^1) (y_4^3) (z_{13})^2 (z_{24})
    -  (y_2^3) (y_4^3) (z_{13})^2 (z_{24}) \nonumber\\&
    + (y_2^1) (y_3^1) (z_{13}) (z_{14}) (z_{24})
    -  (y_2^3) (y_3^1) (z_{13}) (z_{14}) (z_{24})
    + 2 (y_2^1) (y_3^2) (z_{13}) (z_{14}) (z_{24}) \nonumber\\&
    + (y_1^3) (y_4^2) (z_{13}) (z_{23}) (z_{24}) 
    - 2 (y_1^2) (y_4^3) (z_{13}) (z_{23}) (z_{24}) 
    + 2 (y_1^2) (y_3^1) (z_{14}) (z_{23}) (z_{24}) \nonumber\\&
    + (y_1^2) (y_3^2) (z_{14}) (z_{23}) (z_{24}) 
    -  (y_1^3) (y_3^2) (z_{14}) (z_{23}) (z_{24})
    - 2 s (z_{13}) (z_{14}) (z_{23}) (z_{24})\nonumber\\&
    -  (y_1^2) (y_3^1) (z_{13}) (z_{24})^2
    - 2 (y_1^2) (y_3^2) (z_{13}) (z_{24})^2 
    -  (y_1^3) (y_3^2) (z_{13}) (z_{24})^2 \nonumber\\&
    -  \tfrac{\mathsf{t}}{4}  (z_{13})^2 (z_{24})^2 
    + (y_3^1) (y_4^2) (z_{12})^2 (z_{34}) 
    -  (y_3^2) (y_4^2) (z_{12})^2 (z_{34}) \nonumber\\&
    -  (y_3^2) (y_4^3) (z_{12})^2 (z_{34}) 
    + 2 (y_2^3) (y_4^2) (z_{12}) (z_{13}) (z_{34}) 
    + (y_2^1) (y_4^3) (z_{12}) (z_{13}) (z_{34})\nonumber\\&
    + 2 (y_2^3) (y_4^3) (z_{12}) (z_{13}) (z_{34}) 
    + (y_2^1) (y_3^1) (z_{12}) (z_{14}) (z_{34}) 
    + 2 (y_2^3) (y_3^1) (z_{12}) (z_{14}) (z_{34}) \nonumber\\&
    -  (y_2^1) (y_3^2) (z_{12}) (z_{14}) (z_{34})
    - 2 (y_1^3) (y_4^2) (z_{12}) (z_{23}) (z_{34}) 
    + (y_1^2) (y_4^3) (z_{12}) (z_{23}) (z_{34})\nonumber\\&
    + 2 (y_1^3) (y_2^1) (z_{14}) (z_{23}) (z_{34}) 
    -  (y_1^2) (y_2^3) (z_{14}) (z_{23}) (z_{34}) 
    + (y_1^3) (y_2^3) (z_{14}) (z_{23}) (z_{34}) \nonumber\\&
    - 2 \mathsf{t} (z_{12}) (z_{14}) (z_{23}) (z_{34}) 
    -  (y_1^2) (y_3^1) (z_{12}) (z_{24}) (z_{34}) 
    + (y_1^2) (y_3^2) (z_{12}) (z_{24}) (z_{34}) \nonumber\\&
    + 2 (y_1^3) (y_3^2) (z_{12}) (z_{24}) (z_{34})
    -  (y_1^3) (y_2^1) (z_{13}) (z_{24}) (z_{34}) 
    + 2 (y_1^2) (y_2^3) (z_{13}) (z_{24}) (z_{34})\nonumber\\&
    + (y_1^3) (y_2^3) (z_{13}) (z_{24}) (z_{34}) 
    - 2 \mathsf{u} (z_{12}) (z_{13}) (z_{24}) (z_{34}) 
    -  (y_1^3) (y_2^1) (z_{12}) (z_{34})^2\nonumber\\&
    -  (y_1^2) (y_2^3) (z_{12}) (z_{34})^2 
    - 2 (y_1^3) (y_2^3) (z_{12}) (z_{34})^2 
    -  \tfrac{\mathsf{s}}{4} (z_{12})^2 (z_{34})^2\Bigg]\,,
\end{align}}

\noindent together with the following homogeneous solution ${\mathcal{C}^{(4)}}$:

{\small\allowdisplaybreaks
\begin{align}
    \mathcal{C}^{(4)}&=
    - \frac{16  \mathsf{s} \mathsf{t} h_{14}^2 h_{23}^2}{ \mathsf{u}^5} 
    -  \frac{32  \mathsf{s} h_{13} h_{14} h_{23} h_{24}}{t^2  \mathsf{u}^2} 
    -  \frac{16  \mathsf{s}  \mathsf{u} h_{13}^2 h_{24}^2}{t^5} 
    -  \frac{32 \mathsf{t} h_{12} h_{14} h_{23} h_{34}}{ \mathsf{s}^2  \mathsf{u}^2} 
    -  \frac{32  \mathsf{u} h_{12} h_{13} h_{24} h_{34}}{ \mathsf{s}^2 \mathsf{t}^2}\nonumber\\&
    -  \frac{32 h_{14} (h_1^{34}) h_{23} (h_2^{34}) h_{34}}{ \mathsf{s}^2  \mathsf{u}^3}
    -  \frac{32 h_{13} (h_1^{34}) h_{24} (h_2^{34}) h_{34}}{ \mathsf{s}^2 \mathsf{t}^3} 
    + \frac{16  \mathsf{u} ( \mathsf{s} +  \mathsf{u}) h_{12}^2 h_{34}^2}{ \mathsf{s}^5} 
    -  \frac{32 h_{12} (h_1^{34}) (h_2^{34}) h_{34}^2}{ \mathsf{s}^5} \nonumber\\&
    -  \frac{16 (h_1^{34})^2 (h_2^{34})^2 h_{34}^2}{ \mathsf{s}^5 \mathsf{t}  \mathsf{u}} 
    + \frac{32 h_{14} (h_1^{34}) h_{23} h_{24} (h_3^{12})}{t^2  \mathsf{u}^3} 
    + \frac{32 h_{13} (h_1^{34}) h_{24}^2 (h_3^{12})}{t^5} 
    -  \frac{32 h_{14}^2 h_{23} (h_2^{34}) (h_3^{12})}{ \mathsf{u}^5} \nonumber\\&
    -  \frac{32 h_{13} h_{14} h_{24} (h_2^{34}) (h_3^{12})}{t^3  \mathsf{u}^2} 
    + \frac{32 h_{12} (h_1^{34}) h_{24} h_{34} (h_3^{12})}{ \mathsf{s}^3 \mathsf{t}^2} 
    -  \frac{32 h_{12} h_{14} (h_2^{34}) h_{34} (h_3^{12})}{ \mathsf{s}^3  \mathsf{u}^2} 
    + \frac{32 (h_1^{34})^2 h_{24} (h_2^{34}) h_{34} (h_3^{12})}{ \mathsf{s}^3 \mathsf{t}^3  \mathsf{u}}\nonumber\\&
    -  \frac{32 h_{14} (h_1^{34}) (h_2^{34})^2 h_{34} (h_3^{12})}{ \mathsf{s}^3 \mathsf{t}  \mathsf{u}^3}
    -  \frac{16 (h_1^{34})^2 h_{24}^2 (h_3^{12})^2}{ \mathsf{s} \mathsf{t}^5  \mathsf{u}} 
    + \frac{32 h_{14} (h_1^{34}) h_{24} (h_2^{34}) (h_3^{12})^2}{ \mathsf{s} \mathsf{t}^3  \mathsf{u}^3} 
    -  \frac{16 h_{14}^2 (h_2^{34})^2 (h_3^{12})^2}{ \mathsf{s} \mathsf{t}  \mathsf{u}^5}\nonumber\\&
    -  \frac{32 h_{14} (h_1^{34}) h_{23}^2 ( h_4^{12})}{ \mathsf{u}^5} 
    -  \frac{32 h_{13} (h_1^{34}) h_{23} h_{24} ( h_4^{12})}{t^3  \mathsf{u}^2} 
    + \frac{32 h_{13} h_{14} h_{23} (h_2^{34}) ( h_4^{12})}{t^2  \mathsf{u}^3} 
    + \frac{32 h_{13}^2 h_{24} (h_2^{34}) ( h_4^{12})}{t^5}\nonumber\\&
    -  \frac{32 h_{12} (h_1^{34}) h_{23} h_{34} ( h_4^{12})}{ \mathsf{s}^3  \mathsf{u}^2}
    + \frac{32 h_{12} h_{13} (h_2^{34}) h_{34} ( h_4^{12})}{ \mathsf{s}^3 \mathsf{t}^2}
    -  \frac{32 (h_1^{34})^2 h_{23} (h_2^{34}) h_{34} ( h_4^{12})}{ \mathsf{s}^3 \mathsf{t}  \mathsf{u}^3} 
    + \frac{32 h_{13} (h_1^{34}) (h_2^{34})^2 h_{34} ( h_4^{12})}{ \mathsf{s}^3 \mathsf{t}^3  \mathsf{u}} \nonumber\\&
    -  \frac{32 h_{12} h_{14} h_{23} (h_3^{12}) ( h_4^{12})}{ \mathsf{s}^2  \mathsf{u}^3}
    -  \frac{32 h_{12} h_{13} h_{24} (h_3^{12}) ( h_4^{12})}{ \mathsf{s}^2 \mathsf{t}^3} 
    + \frac{32 (h_1^{34})^2 h_{23} h_{24} (h_3^{12}) ( h_4^{12})}{ \mathsf{s} \mathsf{t}^3  \mathsf{u}^3} \nonumber\\&
    + \frac{32 (2  \mathsf{s}^2 + 2  \mathsf{s}  \mathsf{u} 
    +  \mathsf{u}^2) h_{14} (h_1^{34}) h_{23} (h_2^{34}) (h_3^{12}) ( h_4^{12})}{ \mathsf{s}^2 \mathsf{t}^2  \mathsf{u}^5}
    + \frac{32 ( \mathsf{s}^2 +  \mathsf{u}^2) h_{13} (h_1^{34}) h_{24} (h_2^{34}) (h_3^{12}) ( h_4^{12})}{ \mathsf{s}^2 \mathsf{t}^5  \mathsf{u}^2}\nonumber\\&
    + \frac{32 h_{13} h_{14} (h_2^{34})^2 (h_3^{12}) ( h_4^{12})}{ \mathsf{s} \mathsf{t}^3  \mathsf{u}^3} 
    -  \frac{32 h_{12}^2 h_{34} (h_3^{12}) ( h_4^{12})}{ \mathsf{s}^5}
    + \frac{32 ( \mathsf{s}^2 + 2  \mathsf{s}  \mathsf{u} + 2  \mathsf{u}^2) h_{12} (h_1^{34}) (h_2^{34}) h_{34} (h_3^{12}) ( h_4^{12})}{ \mathsf{s}^5 \mathsf{t}^2  \mathsf{u}^2} \nonumber\\&
    + \frac{32 ( \mathsf{s}^2 +  \mathsf{s}  \mathsf{u} +  \mathsf{u}^2) (h_1^{34})^2 (h_2^{34})^2 h_{34} (h_3^{12}) ( h_4^{12})}{ \mathsf{s}^5 \mathsf{t}^3  \mathsf{u}^3} + \frac{32 h_{12} (h_1^{34}) h_{24} (h_3^{12})^2 ( h_4^{12})}{ \mathsf{s}^3 \mathsf{t}^3  \mathsf{u}}\nonumber\\&
    -  \frac{32 h_{12} h_{14} (h_2^{34}) (h_3^{12})^2 ( h_4^{12})}{ \mathsf{s}^3 \mathsf{t}  \mathsf{u}^3}
    -  \frac{32 ( \mathsf{s}^2 +  \mathsf{s}  \mathsf{u} +  \mathsf{u}^2) (h_1^{34})^2 h_{24} (h_2^{34}) (h_3^{12})^2 ( h_4^{12})}{ \mathsf{s}^3 \mathsf{t}^5  \mathsf{u}^3} \nonumber\\&
    + \frac{32 ( \mathsf{s}^2 +  \mathsf{s}  \mathsf{u} +  \mathsf{u}^2) h_{14} (h_1^{34}) (h_2^{34})^2 (h_3^{12})^2 ( h_4^{12})}{ \mathsf{s}^3 \mathsf{t}^3  \mathsf{u}^5} \nonumber\\&
    -  \frac{16 (h_1^{34})^2 h_{23}^2 ( h_4^{12})^2}{ \mathsf{s} \mathsf{t}  \mathsf{u}^5} + \frac{32 h_{13} (h_1^{34}) h_{23} (h_2^{34}) ( h_4^{12})^2}{ \mathsf{s} \mathsf{t}^3  \mathsf{u}^3} 
    -  \frac{16 h_{13}^2 (h_2^{34})^2 ( h_4^{12})^2}{ \mathsf{s} \mathsf{t}^5  \mathsf{u}}\nonumber\\&
    -  \frac{32 h_{12} (h_1^{34}) h_{23} (h_3^{12}) ( h_4^{12})^2}{ \mathsf{s}^3 \mathsf{t}  \mathsf{u}^3} 
    + \frac{32 h_{12} h_{13} (h_2^{34}) (h_3^{12}) ( h_4^{12})^2}{ \mathsf{s}^3 \mathsf{t}^3  \mathsf{u}} 
    + \frac{32 ( \mathsf{s}^2 +  \mathsf{s}  \mathsf{u} +  \mathsf{u}^2) (h_1^{34})^2 h_{23} (h_2^{34}) (h_3^{12}) ( h_4^{12})^2}{ \mathsf{s}^3 \mathsf{t}^3  \mathsf{u}^5}\nonumber\\&
    -  \frac{32 ( \mathsf{s}^2 +  \mathsf{s}  \mathsf{u} +  \mathsf{u}^2) h_{13} (h_1^{34}) (h_2^{34})^2 (h_3^{12}) ( h_4^{12})^2}{ \mathsf{s}^3 \mathsf{t}^5  \mathsf{u}^3} 
    -  \frac{16 h_{12}^2 (h_3^{12})^2 ( h_4^{12})^2}{ \mathsf{s}^5 \mathsf{t}  \mathsf{u}}\nonumber\\& + \frac{32 ( \mathsf{s}^2 +  \mathsf{s}  \mathsf{u} +  \mathsf{u}^2) h_{12} (h_1^{34}) (h_2^{34}) (h_3^{12})^2 ( h_4^{12})^2}{ \mathsf{s}^5 \mathsf{t}^3  \mathsf{u}^3} 
    -  \frac{16 ( \mathsf{s}^2 +  \mathsf{s}  \mathsf{u} +  \mathsf{u}^2)^2 (h_1^{34})^2 (h_2^{34})^2 (h_3^{12})^2 ( h_4^{12})^2}{ \mathsf{s}^5 \mathsf{t}^5  \mathsf{u}^5}\,.
\end{align}}

\noindent Again, the price to pay for manifest gauge invariance is that locality and factorisation are not manifest. 

In conclusion, our formalism can be easily implemented within a computer algebra program and given a set of exchange amplitudes allows to quickly find (if it exists) the corresponding contact term and as well the manifestly gauge invariant form of the homogeneous solution! With our formalism it is for instance straightforward to rule out local spin-3 self-interactions as well as interaction of a spin-3 field with any spectrum of higher-spin fields if a coupling of spin-3 and a massless spin-2 is non-vanishing.

\bibliographystyle{JHEP}
\bibliography{ref}

\end{document}